\documentclass[twocolumn,letterpaper]{aastex61}

\usepackage{graphicx}	
\usepackage{amsmath}	
\usepackage{amssymb}	
\usepackage{comment}
\usepackage{xspace}
\usepackage{lastpage}

\usepackage{eso-pic}

\def\reff@jnl#1{{\rm#1\/}}
\def\aj{\reff@jnl{AJ}}         
\def\araa{\reff@jnl{ARA\&A}}      
\def\apj{\reff@jnl{ApJ}}        
\def\apjl{\reff@jnl{ApJ}}        
\def\apjs{\reff@jnl{ApJS}}       
\def\aap{\reff@jnl{A\&A}}        
\def\aapr{\reff@jnl{A\&A~Rev.}}     
\def\aaps{\reff@jnl{A\&AS}}       
\def\mnras{\reff@jnl{MNRAS}}      
\def\physrep{\reff@jnl{Physics Reports}}
\def\prd{\reff@jnl{Phys.Rev.D}}     
\def\prl{\reff@jnl{Phys.Rev.Lett}}   
\def\pasp{\reff@jnl{PASP}}       
\def\pasj{\reff@jnl{PASJ}}       
\def\nat{\reff@jnl{Nature}}       
\def\jcap{\reff@jnl{JCAP}}   
\def\memsai{\reff@jnl{MemSAI}} 
\def\na{\reff@jnl{New Astronomy}}       
\def\procspie{\reff@jnl{SPIE}}       
\def\pasa{\reff@jnl{PASA}}       

\def\Sref#1{$\S$\ref{#1}\xspace}

\def\Fref#1{Fig.~\ref{#1}\xspace}

\def\Tref#1{Table~\ref{#1}\xspace}

\def\Eref#1{Eq.~(\ref{#1})\xspace}

\def\Aref#1{Appendix~\ref{#1}\xspace}
\def\Cref#1{Chapter~\ref{#1}\xspace}

\defcitealias{Diemer2014}{DK14}
\defcitealias{Adhikari2014}{A14}
\defcitealias{More2016}{M16}
\defcitealias{Adhikari2016}{A16}
\defcitealias{Baxter2017}{B17}

\def\redmapper{\textsc{redMaPPer}\xspace}

\begin{document}

\title{The Splashback Feature around DES Galaxy Clusters: \\
Galaxy Density and Weak Lensing Profiles}

\correspondingauthor{Chihway Chang}
\email{chihway@kicp.uchicago.edu}

\author{C.~Chang}
\affiliation{Kavli Institute for Cosmological Physics, University of Chicago, Chicago, IL 60637, USA}
\author{E.~Baxter}
\affiliation{Department of Physics and Astronomy, University of Pennsylvania, Philadelphia, PA 19104, USA}
\author{B.~Jain}
\affiliation{Department of Physics and Astronomy, University of Pennsylvania, Philadelphia, PA 19104, USA}
\author{C.~S{\'a}nchez}
\affiliation{Department of Physics and Astronomy, University of Pennsylvania, Philadelphia, PA 19104, USA}
\affiliation{Institut de F\'{\i}sica d'Altes Energies (IFAE), The Barcelona Institute of Science and Technology, Campus UAB, 08193 Bellaterra (Barcelona) Spain}
\author{S.~Adhikari}
\affiliation{Department of Astronomy, University of Illinois at Urbana-Champaign, Champaign, IL 61801, USA}
\affiliation{Kavli Institute for Particle Astrophysics \& Cosmology, P. O. Box 2450, Stanford University, Stanford, CA 94305, USA}
\author{T.~N.~Varga}
\affiliation{Universit\"ats-Sternwarte, Fakult\"at f\"ur Physik, Ludwig-Maximilians Universit\"at M\"unchen, Scheinerstr. 1, 81679 M\"unchen, Germany}
\affiliation{Max Planck Institute for Extraterrestrial Physics, Giessenbachstrasse, 85748 Garching, Germany}
\author{Y.~Fang}
\affiliation{Department of Physics and Astronomy, University of Pennsylvania, Philadelphia, PA 19104, USA}
\author{E.~Rozo}
\affil{Department of Physics, University of Arizona, Tucson, AZ 85721, USA}
\author{E.~S.~Rykoff}
\affiliation{Kavli Institute for Particle Astrophysics \& Cosmology, P. O. Box 2450, Stanford University, Stanford, CA 94305, USA}
\affiliation{SLAC National Accelerator Laboratory, Menlo Park, CA 94025, USA}
\author{A.~Kravtsov}
\affil{Kavli Institute for Cosmological Physics, The University of Chicago, Chicago, IL 60637, USA}
\affil{Department of Astronomy and Astrophysics, The University of Chicago, Chicago, IL 60637, USA}
\affil{Enrico Fermi Institute, The University of Chicago, Chicago, IL 60637, USA}
\author{D.~Gruen}
\affiliation{Kavli Institute for Particle Astrophysics \& Cosmology, P. O. Box 2450, Stanford University, Stanford, CA 94305, USA}
\affiliation{SLAC National Accelerator Laboratory, Menlo Park, CA 94025, USA}
\author{W.~Hartley}
\affiliation{Department of Physics \& Astronomy, University College London, Gower Street, London, WC1E 6BT, UK}
\author{E.~M.~Huff}
\affiliation{Jet Propulsion Laboratory, California Institute of Technology, 4800 Oak Grove Dr., Pasadena, CA 91109, USA}
\author{M.~Jarvis}
\affiliation{Department of Physics and Astronomy, University of Pennsylvania, Philadelphia, PA 19104, USA}
\author{A.~G.~Kim}
\affiliation{Lawrence Berkeley National Laboratory, 1 Cyclotron Road, Berkeley, CA 94720, USA}
\author{J.~Prat}
\affiliation{Institut de F\'{\i}sica d'Altes Energies (IFAE), The Barcelona Institute of Science and Technology, Campus UAB, 08193 Bellaterra (Barcelona) Spain}
\author{N.~MacCrann}
\affiliation{Center for Cosmology and Astro-Particle Physics, The Ohio State University, Columbus, OH 43210, USA}
\affiliation{Department of Physics, The Ohio State University, Columbus, OH 43210, USA}
\author{T.~McClintock}
\affiliation{Department of Physics, University of Arizona, Tucson, AZ 85721, USA}
\author{A.~Palmese}
\affiliation{Department of Physics \& Astronomy, University College London, Gower Street, London, WC1E 6BT, UK}
\author{D.~Rapetti}
\affiliation{Center for Astrophysics and Space Astronomy, Department of Astrophysical and Planetary Science, University of Colorado, Boulder, C0 80309, USA}
\affiliation{NASA Ames Research Center, Moffett Field, CA 94035, USA}
\author{R.~P.~Rollins}
\affiliation{Jodrell Bank Center for Astrophysics, School of Physics and Astronomy, University of Manchester, Oxford Road, Manchester, M13 9PL, UK}
\author{S.~Samuroff}
\affiliation{Jodrell Bank Center for Astrophysics, School of Physics and Astronomy, University of Manchester, Oxford Road, Manchester, M13 9PL, UK}
\author{E.~Sheldon}
\affiliation{Brookhaven National Laboratory, Bldg 510, Upton, NY 11973, USA}
\author{M.~A.~Troxel}
\affiliation{Center for Cosmology and Astro-Particle Physics, The Ohio State University, Columbus, OH 43210, USA}
\affiliation{Department of Physics, The Ohio State University, Columbus, OH 43210, USA}
\author{R.~H.~Wechsler}
\affiliation{Kavli Institute for Particle Astrophysics \& Cosmology, P. O. Box 2450, Stanford University, Stanford, CA 94305, USA}
\affiliation{SLAC National Accelerator Laboratory, Menlo Park, CA 94025, USA}
\affiliation{Department of Physics, Stanford University, 382 Via Pueblo Mall, Stanford, CA 94305, USA}
\author{Y.~Zhang}
\affiliation{Fermi National Accelerator Laboratory, P. O. Box 500, Batavia, IL 60510, USA}
\author{J.~Zuntz}
\affiliation{Institute for Astronomy, University of Edinburgh, Edinburgh EH9 3HJ, UK}
\author{T.~M.~C.~Abbott}
\affiliation{Cerro Tololo Inter-American Observatory, National Optical Astronomy Observatory, Casilla 603, La Serena, Chile}
\author{F.~B.~Abdalla}
\affiliation{Department of Physics \& Astronomy, University College London, Gower Street, London, WC1E 6BT, UK}
\affiliation{Department of Physics and Electronics, Rhodes University, PO Box 94, Grahamstown, 6140, South Africa}
\author{S.~Allam}
\affiliation{Fermi National Accelerator Laboratory, P. O. Box 500, Batavia, IL 60510, USA}
\author{J.~Annis}
\affiliation{Fermi National Accelerator Laboratory, P. O. Box 500, Batavia, IL 60510, USA}
\author{K.~Bechtol}
\affiliation{LSST, 933 North Cherry Avenue, Tucson, AZ 85721, USA}
\author{A.~Benoit-L{\'e}vy}
\affiliation{Department of Physics \& Astronomy, University College London, Gower Street, London, WC1E 6BT, UK}
\affiliation{CNRS, UMR 7095, Institut d'Astrophysique de Paris, F-75014, Paris, France}
\affiliation{Sorbonne Universit\'es, UPMC Univ Paris 06, UMR 7095, Institut d'Astrophysique de Paris, F-75014, Paris, France}
\author{G.~M.~Bernstein}
\affiliation{Department of Physics and Astronomy, University of Pennsylvania, Philadelphia, PA 19104, USA}
\author{D.~Brooks}
\affiliation{Department of Physics \& Astronomy, University College London, Gower Street, London, WC1E 6BT, UK}
\author{E.~Buckley-Geer}
\affiliation{Fermi National Accelerator Laboratory, P. O. Box 500, Batavia, IL 60510, USA}
\author{A.~Carnero~Rosell}
\affiliation{Laborat\'orio Interinstitucional de e-Astronomia - LIneA, Rua Gal. Jos\'e Cristino 77, Rio de Janeiro, RJ - 20921-400, Brazil}
\affiliation{Observat\'orio Nacional, Rua Gal. Jos\'e Cristino 77, Rio de Janeiro, RJ - 20921-400, Brazil}
\author{M.~Carrasco~Kind}
\affiliation{Department of Astronomy, University of Illinois, 1002 W. Green Street, Urbana, IL 61801, USA}
\affiliation{National Center for Supercomputing Applications, 1205 West Clark St., Urbana, IL 61801, USA}
\author{J.~Carretero}
\affiliation{Institut de F\'{\i}sica d'Altes Energies (IFAE), The Barcelona Institute of Science and Technology, Campus UAB, 08193 Bellaterra (Barcelona) Spain}
\author{C.~B.~D'Andrea}
\affiliation{Department of Physics and Astronomy, University of Pennsylvania, Philadelphia, PA 19104, USA}
\author{L.~N.~da Costa}
\affiliation{Laborat\'orio Interinstitucional de e-Astronomia - LIneA, Rua Gal. Jos\'e Cristino 77, Rio de Janeiro, RJ - 20921-400, Brazil}
\affiliation{Observat\'orio Nacional, Rua Gal. Jos\'e Cristino 77, Rio de Janeiro, RJ - 20921-400, Brazil}
\author{C.~Davis}
\affiliation{Kavli Institute for Particle Astrophysics \& Cosmology, P. O. Box 2450, Stanford University, Stanford, CA 94305, USA}
\author{S.~Desai}
\affiliation{Department of Physics, IIT Hyderabad, Kandi, Telangana 502285, India}
\author{H.~T.~Diehl}
\affiliation{Fermi National Accelerator Laboratory, P. O. Box 500, Batavia, IL 60510, USA}
\author{J.~P.~Dietrich}
\affiliation{Excellence Cluster Universe, Boltzmannstr.\ 2, 85748 Garching, Germany}
\affiliation{Faculty of Physics, Ludwig-Maximilians-Universit\"at, Scheinerstr. 1, 81679 Munich, Germany}
\author{A.~Drlica-Wagner}
\affiliation{Fermi National Accelerator Laboratory, P. O. Box 500, Batavia, IL 60510, USA}
\author{T.~F.~Eifler}
\affiliation{Jet Propulsion Laboratory, California Institute of Technology, 4800 Oak Grove Dr., Pasadena, CA 91109, USA}
\affiliation{Department of Physics, California Institute of Technology, Pasadena, CA 91125, USA}
\author{B.~Flaugher}
\affiliation{Fermi National Accelerator Laboratory, P. O. Box 500, Batavia, IL 60510, USA}
\author{P.~Fosalba}
\affiliation{Institute of Space Sciences, IEEC-CSIC, Campus UAB, Carrer de Can Magrans, s/n,  08193 Barcelona, Spain}
\author{J.~Frieman}
\affiliation{Kavli Institute for Cosmological Physics, University of Chicago, Chicago, IL 60637, USA}
\affiliation{Fermi National Accelerator Laboratory, P. O. Box 500, Batavia, IL 60510, USA}
\author{J.~Garc\'ia-Bellido}
\affiliation{Instituto de Fisica Teorica UAM/CSIC, Universidad Autonoma de Madrid, 28049 Madrid, Spain}
\author{E.~Gaztanaga}
\affiliation{Institute of Space Sciences, IEEC-CSIC, Campus UAB, Carrer de Can Magrans, s/n,  08193 Barcelona, Spain}
\author{D.~W.~Gerdes}
\affiliation{Department of Astronomy, University of Michigan, Ann Arbor, MI 48109, USA}
\affiliation{Department of Physics, University of Michigan, Ann Arbor, MI 48109, USA}
\author{R.~A.~Gruendl}
\affiliation{Department of Astronomy, University of Illinois, 1002 W. Green Street, Urbana, IL 61801, USA}
\affiliation{National Center for Supercomputing Applications, 1205 West Clark St., Urbana, IL 61801, USA}
\author{J.~Gschwend}
\affiliation{Laborat\'orio Interinstitucional de e-Astronomia - LIneA, Rua Gal. Jos\'e Cristino 77, Rio de Janeiro, RJ - 20921-400, Brazil}
\affiliation{Observat\'orio Nacional, Rua Gal. Jos\'e Cristino 77, Rio de Janeiro, RJ - 20921-400, Brazil}
\author{G.~Gutierrez}
\affiliation{Fermi National Accelerator Laboratory, P. O. Box 500, Batavia, IL 60510, USA}
\author{K.~Honscheid}
\affiliation{Center for Cosmology and Astro-Particle Physics, The Ohio State University, Columbus, OH 43210, USA}
\affiliation{Department of Physics, The Ohio State University, Columbus, OH 43210, USA}
\author{D.~J.~James}
\affiliation{Astronomy Department, University of Washington, Box 351580, Seattle, WA 98195, USA}
\author{T.~Jeltema}
\affiliation{Santa Cruz Institute for Particle Physics, Santa Cruz, CA 95064, USA}
\author{E.~Krause}
\affiliation{Kavli Institute for Particle Astrophysics \& Cosmology, P. O. Box 2450, Stanford University, Stanford, CA 94305, USA}
\author{K.~Kuehn}
\affiliation{Australian Astronomical Observatory, North Ryde, NSW 2113, Australia}
\author{O.~Lahav}
\affiliation{Department of Physics \& Astronomy, University College London, Gower Street, London, WC1E 6BT, UK}
\author{M.~Lima}
\affiliation{Laborat\'orio Interinstitucional de e-Astronomia - LIneA, Rua Gal. Jos\'e Cristino 77, Rio de Janeiro, RJ - 20921-400, Brazil}
\affiliation{Departamento de F\'isica Matem\'atica, Instituto de F\'isica, Universidade de S\~ao Paulo, CP 66318, S\~ao Paulo, SP, 05314-970, Brazil}
\author{M.~March}
\affiliation{Department of Physics and Astronomy, University of Pennsylvania, Philadelphia, PA 19104, USA}
\author{J.~L.~Marshall}
\affiliation{George P. and Cynthia Woods Mitchell Institute for Fundamental Physics and Astronomy, and Department of Physics and Astronomy, Texas A\&M University, College Station, TX 77843,  USA}
\author{P.~Martini}
\affiliation{Center for Cosmology and Astro-Particle Physics, The Ohio State University, Columbus, OH 43210, USA}
\affiliation{Department of Astronomy, The Ohio State University, Columbus, OH 43210, USA}
\author{P.~Melchior}
\affiliation{Department of Astrophysical Sciences, Princeton University, Peyton Hall, Princeton, NJ 08544, USA}
\author{F.~Menanteau}
\affiliation{Department of Astronomy, University of Illinois, 1002 W. Green Street, Urbana, IL 61801, USA}
\affiliation{National Center for Supercomputing Applications, 1205 West Clark St., Urbana, IL 61801, USA}
\author{R.~Miquel}
\affiliation{Institut de F\'{\i}sica d'Altes Energies (IFAE), The Barcelona Institute of Science and Technology, Campus UAB, 08193 Bellaterra (Barcelona) Spain}
\affiliation{Instituci\'o Catalana de Recerca i Estudis Avan\c{c}ats, E-08010 Barcelona, Spain}
\author{J.~J.~Mohr}
\affiliation{Max Planck Institute for Extraterrestrial Physics, Giessenbachstrasse, 85748 Garching, Germany}
\affiliation{Excellence Cluster Universe, Boltzmannstr.\ 2, 85748 Garching, Germany}
\affiliation{Faculty of Physics, Ludwig-Maximilians-Universit\"at, Scheinerstr. 1, 81679 Munich, Germany}
\author{B.~Nord}
\affiliation{Fermi National Accelerator Laboratory, P. O. Box 500, Batavia, IL 60510, USA}
\author{R.~L.~C.~Ogando}
\affiliation{Laborat\'orio Interinstitucional de e-Astronomia - LIneA, Rua Gal. Jos\'e Cristino 77, Rio de Janeiro, RJ - 20921-400, Brazil}
\affiliation{Observat\'orio Nacional, Rua Gal. Jos\'e Cristino 77, Rio de Janeiro, RJ - 20921-400, Brazil}
\author{A.~A.~Plazas}
\affiliation{Jet Propulsion Laboratory, California Institute of Technology, 4800 Oak Grove Dr., Pasadena, CA 91109, USA}
\author{E.~Sanchez}
\affiliation{Centro de Investigaciones Energ\'eticas, Medioambientales y Tecnol\'ogicas (CIEMAT), Madrid, Spain}
\author{V.~Scarpine}
\affiliation{Fermi National Accelerator Laboratory, P. O. Box 500, Batavia, IL 60510, USA}
\author{R.~Schindler}
\affiliation{SLAC National Accelerator Laboratory, Menlo Park, CA 94025, USA}
\author{M.~Schubnell}
\affiliation{Department of Physics, University of Michigan, Ann Arbor, MI 48109, USA}
\author{I.~Sevilla-Noarbe}
\affiliation{Centro de Investigaciones Energ\'eticas, Medioambientales y Tecnol\'ogicas (CIEMAT), Madrid, Spain}
\author{M.~Smith}
\affiliation{School of Physics and Astronomy, University of Southampton,  Southampton, SO17 1BJ, UK}
\author{R.~C.~Smith}
\affiliation{Cerro Tololo Inter-American Observatory, National Optical Astronomy Observatory, Casilla 603, La Serena, Chile}
\author{M.~Soares-Santos}
\affiliation{Fermi National Accelerator Laboratory, P. O. Box 500, Batavia, IL 60510, USA}
\author{F.~Sobreira}
\affiliation{Laborat\'orio Interinstitucional de e-Astronomia - LIneA, Rua Gal. Jos\'e Cristino 77, Rio de Janeiro, RJ - 20921-400, Brazil}
\affiliation{Instituto de F\'isica Gleb Wataghin, Universidade Estadual de Campinas, 13083-859, Campinas, SP, Brazil}
\author{E.~Suchyta}
\affiliation{Computer Science and Mathematics Division, Oak Ridge National Laboratory, Oak Ridge, TN 37831}
\author{M.~E.~C.~Swanson}
\affiliation{National Center for Supercomputing Applications, 1205 West Clark St., Urbana, IL 61801, USA}
\author{G.~Tarle}
\affiliation{Department of Physics, University of Michigan, Ann Arbor, MI 48109, USA}
\author{J.~Weller}
\affiliation{Universit\"ats-Sternwarte, Fakult\"at f\"ur Physik, Ludwig-Maximilians Universit\"at M\"unchen, Scheinerstr. 1, 81679 M\"unchen, Germany}
\affiliation{Max Planck Institute for Extraterrestrial Physics, Giessenbachstrasse, 85748 Garching, Germany}
\affiliation{Excellence Cluster Universe, Boltzmannstr.\ 2, 85748 Garching, Germany}

\collaboration{(DES Collaboration)}
\shorttitle{Splashback in DES}



\begin{abstract}
Splashback refers to the process of matter that is accreting onto a dark matter halo reaching 
its first orbital apocenter and turning around in its orbit. The cluster-centric radius at which this 
process occurs, $r_{\rm sp}$, defines a halo boundary that is connected to the dynamics of the 
cluster. A rapid decline in the halo profile is expected near $r_{\rm sp}$. We measure the 
galaxy number density and weak lensing mass profiles around \redmapper galaxy clusters in the 
first year Dark Energy Survey (DES) data. For a cluster sample with mean $M_{\rm 200m}$ mass 
$\approx 2.5\times10^{14}\,M_{\odot}$, we find strong evidence of a splashback-like steepening of 
the galaxy density profile and measure $r_{\rm sp}=1.13\pm 0.07$ $h^{-1}$Mpc, consistent with 
earlier SDSS measurements of \citet{More2016} and \citet{Baxter2017}. Moreover, our weak 
lensing measurement demonstrates for the first time the existence of a splashback-like steepening 
of the matter profile of galaxy clusters. We measure $r_{\rm sp}=1.34 \pm 0.21$ $h^{-1}$Mpc from 
the weak lensing data, in good agreement with our galaxy density measurements. 
For different cluster and galaxy samples, we find that consistent with $\Lambda$CDM simulations, 
$r_{\rm sp}$ scales with $R_{\rm 200m}$ and does not evolve with redshift over the redshift range 
of 0.3--0.6. We also find that potential systematic effects associated 
with the \redmapper algorithm may impact the location of $r_{\rm sp}$. We discuss progress 
needed to understand the systematic uncertainties and fully exploit forthcoming data from DES and 
future surveys, emphasizing the importance of more realistic mock catalogs and independent cluster samples.

\end{abstract}


\section{Introduction}
\label{sec:intro}

The density profiles of dark matter halos in N-body simulations exhibit a steepening at 
radii comparable to the halo virial radius \citep[][hereafter \citetalias{Diemer2014}]{Diemer2014}. 
Such a feature was predicted by analytical collapse models of \citet{Gunn1972}, 
\citet{Fillmore1984} and \citet{Bertschinger1985}. The sharp decline in the profile can be 
understood as resulting from an absence of particles orbiting beyond the radius of second 
turnaround\footnote{The radius of first turnaround is the radius at which a particle first 
separates from the Hubble flow and begins to fall towards an overdensity. The radius of 
second turnaround is the radius at which a particle that has passed by the halo once turns 
around in its orbit.}. In 
simulations, a phase-space caustic cleanly separates matter that is experiencing second 
turnaround from matter that is on first infall, leading to a very sharp steepening in the halo 
profile \citep[\citetalias{Diemer2014},][]{Adhikari2014, Diemer2017}. As measurements from 
individual halos are noisy, to detect this sharp steepening one needs to ``stack'', or average, 
over a large number of halos. This makes the caustic structure less clear since the dark matter 
halos are oftentimes non-spherical. Nevertheless \citetalias{Diemer2014} showed that some of 
the steepening can in principle be detected based on simulations.
\citet{Mansfield2016} later found that one can improve on the stacking procedure by accounting 
for the effects of subhalos, which sharpens the steepening of the profile even more. This 
feature --- which appears as a narrow minimum in the logarithmic derivative of the halo 
density profile --- has been termed {\it splashback}.

The splashback feature is potentially interesting for several reasons. First, it defines a physical 
boundary of a dark matter halo that is motivated by dynamics \citep{More2015}. This is different 
from other common halo boundary definitions, such as $R_{200m}$ (the radius within which 
the mean density is 200 times the mean density of the Universe at that redshift), which need not 
be associated with any change in physical properties across the boundary. Furthermore, the location 
of the splashback feature has been shown in simulations to correlate with the halo accretion rate 
\citep[\citetalias{Diemer2014},][]{Diemer2017}. Since the feature is in principle straightforward 
to measure in data, it could potentially be used to constrain halo accretion rates of clusters, which 
are otherwise challenging to measure. Finally, the sharpness of the feature and the relatively simple 
dynamics that are responsible for its generation make it a potentially powerful probe of new physics, 
such as dark matter self-interaction.

The first measurement of the splashback feature in data was performed by \citet[][hereafter 
\citetalias{More2016}]{More2016} using DR8 data from the Sloan Digital Sky Survey (SDSS)
\citep{Aihara2011}. \citetalias{More2016} measured the projected galaxy density profiles, 
$\Sigma_g(R)$, around galaxy clusters in the \redmapper catalog of \citet{Rykoff2014}, finding 
evidence for a sharp minimum in the logarithmic derivatives of these profiles. Note, however, that 
a minimum in the logarithmic derivative does not by itself constitute evidence for a splashback 
feature. Indeed, if the matter profile of the halo is described by a Navarro-Frenk-White 
\citep[NFW][]{Navarro1996} profile at small scales and the halo-matter correlation function at 
large scales, there will necessarily be a minimum in the logarithmic derivative in the transition 
regime.  In the language of the halo model \citep[for a review see][]{Cooray:2002}, a minimum of the 
logarithmic derivative is naturally associated with the transition regime between the one-halo and 
the two-halo term. Defining the splashback feature is a way to isolate this dynamical feature 
that is not explicitly described by either the one-halo or the two-halo term. The splashback 
process produces a profile that is significantly steeper at this transition region than what is 
expected from the above naive picture of an NFW profile plus the halo-matter correlation function.

By fitting different models to the measured $\Sigma_g$ profiles, \citetalias{More2016} determined 
that the data show strong evidence of the existence of a splashback feature. However, 
\citetalias{More2016} determined that the location of the splashback feature (henceforth the 
{\it splashback radius}) measured in SDSS data appears to be smaller than that predicted by 
dark matter-only N-body simulations. The explanation for this discrepancy remains unclear; 
\citetalias{More2016} considered several possibilities, including dark matter self-interaction. 
Two followup studies \citep{Zu2016, Busch2017} examined potential systematic effects in the 
estimation of the splashback radius, and showed that projection along the line of sight could 
affect the estimated splashback radius, especially when employing a selection based on 
$\langle R_{\rm mem} \rangle$ (the weighted average member distance to the cluster center; see 
more discussion in \Sref{sec:model_sys}), as was done in \citetalias{More2016}.

On the other hand, \citet[][hereafter \citetalias{Baxter2017}]{Baxter2017} pointed out a difficulty 
associated with quantifying the evidence for a splashback feature using the model parametrizations 
of \citetalias{Diemer2014} and \citetalias{More2016}. In particular, these parametrizations rely on a 
truncated Einasto profile \citep{Einasto1965}, with the truncation term representing the splashback 
feature. However, the Einasto model is sufficiently flexible that even without such truncation, it can 
still reproduce a splashback-like steepening in the outer halo profile. Consequently, the evidence 
for splashback quantified either with a $\Delta \chi^2$ (as in \citealt{More2016}) or a Bayesian 
evidence ratio (as in \citetalias{Baxter2017}) can be misleadingly low, even when there is significant 
steeping of the outer halo profile. This problem becomes more severe when additional flexibility 
is introduced to the model to account for halo miscentering for example.

A more robust approach proposed by \citetalias{Baxter2017} is to instead use the model fits to 
separate the contributions to the total profile from infalling and collapsed material. The logarithmic 
derivative of the profile of the collapsed material can then be used to identify the presence of a 
splashback feature in the density profile. SDSS clusters show a dramatic steepening of the 
collapsed material profile slightly outside the virial radius, consistent with the presence of a splashback 
feature. \citetalias{Baxter2017} also measured the profiles of red and blue galaxies around \redmapper 
clusters, showing that the fraction consisting of red galaxies exhibits a sharp transition at the splashback 
radius. This is consistent with the interpretation of star formation being quenched in galaxies that have 
orbited through the cluster, and adds additional support to the picture of a physically motivated halo 
boundary.
 
Measurement of the splashback radius has also been used recently by \citet[][hereafter 
\citetalias{Adhikari2016}]{Adhikari2016} to measure dynamical friction in galaxy clusters. Dynamical friction 
refers to an effective drag force induced on a massive object via gravitational interaction with nearby matter 
\citep{Chandrasekhar1949, Binney2008}. As pointed out by \citetalias{More2016}, dynamical friction will act 
to reduce the splashback radii of galaxy clusters since a subhalo that has experienced dynamical friction will 
turn around after first infall sooner compared to a subhalo not experiencing dynamical friction. 
\citetalias{Adhikari2016} tested this hypothesis using a sample of lower richness clusters from SDSS, since 
the impact of dynamical friction on a fixed galaxy sample is expected to be larger for low-mass parent halos. 
As the effect of dynamical friction increases with subhalo mass, one expects more massive subhalos to have 
smaller splashback radii. Indeed, \citetalias{Adhikari2016} found that the splashback radius identified 
using a bright galaxy sample (which are expected to live in more massive subhalos) was smaller than the 
splashback radius identified using a fainter galaxy sample. Two caveats to this simple picture are that 
galaxies of different magnitudes may have different orbits through the cluster and that quenching may result 
in changes to galaxy magnitudes within the cluster.

In this work, we measure the galaxy density and weak lensing mass profiles around galaxy clusters 
in data from the first year of the Dark Energy Survey (DES Y1). We carry out analyses based on the 
methodology developed in \citetalias{Baxter2017} to characterize the splashback feature. This new 
data set provides several advances over previous measurements. First, the DES footprint maps a 
different part of the sky from that of SDSS, providing an independent measurement of a non-overlapping 
sample. Second, the DES data extend to higher redshift ranges and fainter galaxies than SDSS. Finally, 
the DES Y1 weak lensing measurements have significantly higher signal-to-noise compared to the previous 
lensing measurements in \citet{Umetsu2017}, who were only able to place a lower bound on the 
location of the splashback radius. While the signal-to-noise of the lensing measurements is still lower 
than that of the galaxy density measurements, lensing has the advantage of directly probing the mass 
profile of the halos in contrast to the galaxy density profile, which makes it cleaner to compare with dark 
matter simulations.

The paper is organized as follows. In \Sref{sec:data}, we introduce the dataset used in this work and 
the selection criteria of the samples. In \Sref{sec:model} we outline the model that is used to describe 
the observed galaxy and weak lensing profiles around clusters. In \Sref{sec:measurements}, the 
measurement methods for both the galaxy and the lensing profiles are described, followed by the model-fitting 
procedure and a summary of the model priors. In \Sref{sec:results} we present the fiducial measurements 
of the galaxy and lensing profiles, as well as comparison with simulations. We then investigate in 
\Sref{sec:df} the topic of dynamical friction by looking at the splashback feature for galaxy samples 
of different luminosities. We discuss the potential systematic effects in the measurements in 
\Sref{sec:model_sys} and conclude in \Sref{sec:discussion}.

If not specified otherwise, throughout the paper, we assume a flat $\rm \Lambda$CDM cosmology with $h=0.7$, 
$\Omega_{m}=0.3$. In addition, all calculations and plots use comoving coordinates.

\section{Data}
\label{sec:data}

The measurements in this work are based on the DES Y1 data
\citep{Diehl2014SPIE}. Here we describe briefly the relevant catalogs
used, including the \redmapper galaxy cluster catalog, the photometric
galaxy catalog, the weak lensing shear catalogs and the photometric
redshift (photo-$z$) catalog.

\subsection{The \redmapper Galaxy Cluster Catalog}

We use a galaxy cluster catalog constructed using the \redmapper
algorithm described in \citet{Rykoff2014, Rykoff2016}. \redmapper is a
red-sequence cluster finder that is optimized for large-scale optical
surveys, such as DES. The same algorithm was employed to construct the
SDSS-based cluster catalog used in \citetalias{More2016} and
\citetalias{Baxter2017}.

\begin{figure}
\centering
\includegraphics[width=\linewidth]{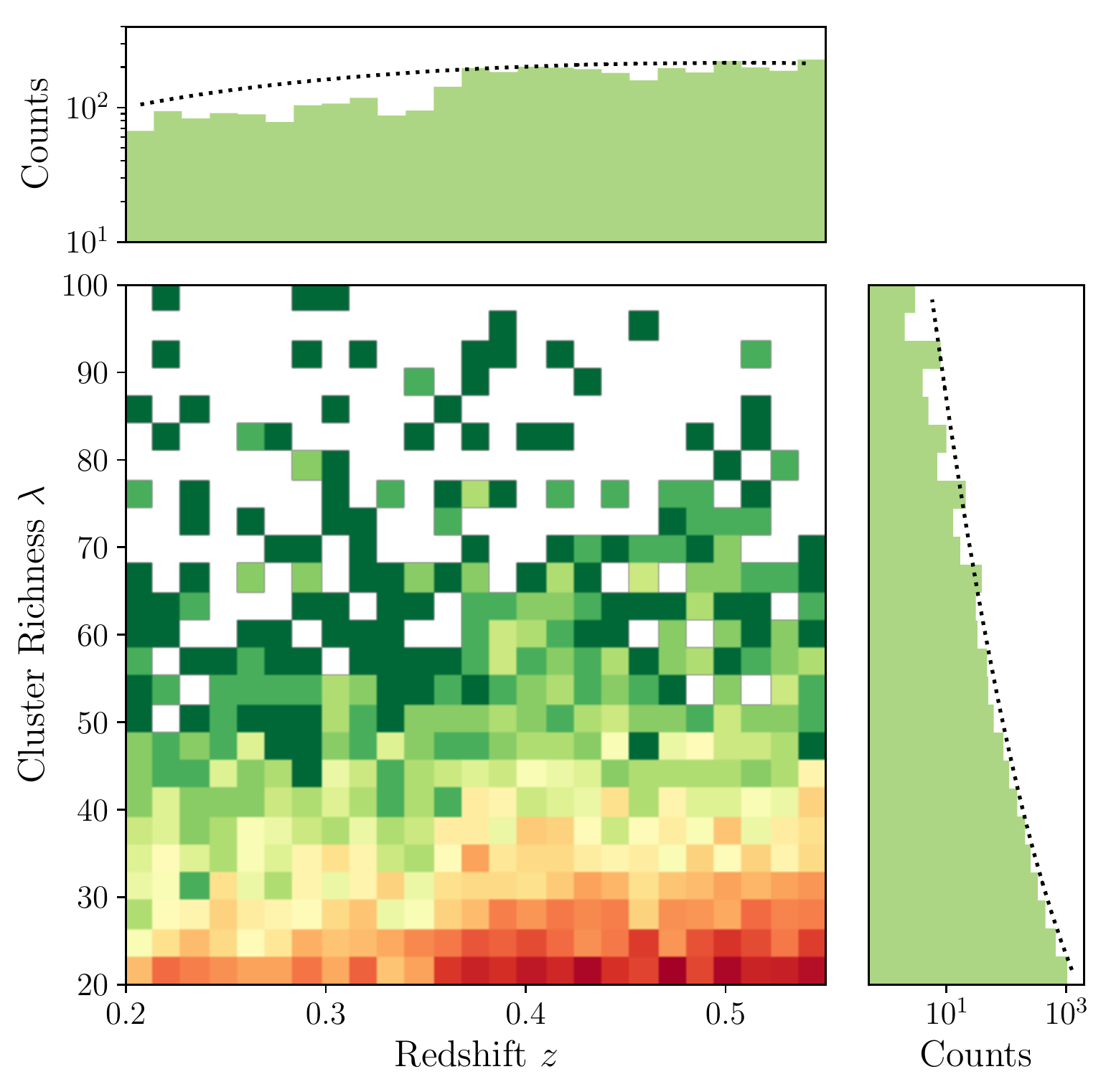}
\caption{The $\lambda$ and $z$ distributions of the fiducial
  \redmapper sample used in this work. The lower left panel shows the
  2D histogram in the $z$-$\lambda$ plane while the upper left and the
  lower right panels show the individual 1D histograms of $z$ and
  $\lambda$. The color bars in the 2D histogram are shown in log
  scale. The naive theoretical expectation of the redshift and
  richness distribution for our sample based on \citet{Tinker2008} and
  \citet{Melchior2016} is shown by the black dashed curves.}
\label{fig:distribution}
\end{figure}

The ``fiducial'' cluster catalog used in this work is constructed from
a volume-limited sample \citep[similar to that described
  in][]{Rykoff2016} with a redshift selection of $0.2<z<0.55$ and a
richness selection of $20<\lambda<100$. We only use the Y1 region at
Dec$<-35^{\circ}$ where most of the tests for the weak lensing and
photo-$z$ catalogs were conducted. The cluster redshifts used for
selection are determined by the \redmapper algorithm and are expected
to have mean uncertainties $\sigma_z \sim 0.01(1+z)$
\citep{Rykoff2016}. We also consider different subsets of the fiducial
sample as well as a ``high-z'' sample that is not contained in the
fiducial sample.
The characteristics of all the samples used in this paper are
summarized in \Tref{tab:clust_sample}, while the redshift-richness
distribution of the fiducial sample is shown in
\Fref{fig:distribution}. The mean mass for each sample listed in
\Tref{tab:clust_sample} is calculated from the mass-richness relation
derived in \citet{Melchior2016} for \redmapper clusters identified in
DES Science Verification (SV) data.  The \redmapper algorithm is
expected to be approximately survey-independent.  However, small
differences the \citet{Melchior2016} mass-richness relation for SV
data and the mass-richness relation of Y1 data may exist due to
e.g. differences in data quality or statistical fluctuations. We rely
on the SV mass calibration here because it was derived using DES data;
the mass calibration of Y1 \redmapper clusters using galaxy lensing is
forthcoming. As further support for our use of the SV mass-richness
relation from \citet{Melchior2016}, we note that \citet{Baxter2017b}
performed a mass calibration of DES Y1 \redmapper clusters using
gravitational lensing of the cosmic background radiation, finding
excellent consistency with the \citet{Melchior2016} results.  We
  also list in \Tref{tab:clust_sample} the mean $R_{\lambda} =
  (\lambda/100)^{0.2} \, h^{-1} {\rm Mpc}$ values in physical units,
  which is used later in modelling the cluster miscentering
  (\Sref{subsec:mis_center}).

We see that the number of clusters falls steeply with richness and
increases by a factor of $\sim2$ over the redshift range. The
structure in the redshift distribution is associated with the DES
filter transition and the 4000 $\buildrel _\circ \over {\mathrm{A}}$
break, where the photo-$z$ redshifts are less certain.  We overlay the
histogram with the theoretical expectation of the number of halos
given mass and redshift using the halo mass function from
\citet{Tinker2008} and the mass-richness relation from
\citet{Melchior2016}.  The data roughly follow the expectation, with
higher discrepancies in the low-richness bins. We also make use of the
\redmapper random catalogs, which uniformly sample the volume over
which a real cluster could have been observed. As described in
\Sref{sec:measurements}, the random catalog is used to estimate the
background mean galaxy distribution in the absence of galaxy clusters.

Finally, uncertainty in the cluster center position is important in
this analysis. According to \citet{Rykoff2016} roughly 22\% of the
clusters are mis-centered at about 0.3$R_{\lambda}$.  We discuss in
\Sref{subsec:mis_center} how this is incorporated into our model.

\subsection{The Photometric Galaxy Catalog}

We use photometric galaxies from the DES Y1 Gold catalog described in
\citet{DrlicaWagner2017}. Our galaxy selection begins with a
flux-limited sample of $i<21.5$ with the following flag cuts:
\texttt{flags\_badregion = flags\_gold = 0}, the following color cuts:
\texttt{-1$<$mag\_auto\_g - mag\_auto\_r$<$3},
\texttt{-1$<$mag\_auto\_r - mag\_auto\_i$<$2.5},
\texttt{-1$<$mag\_auto\_i - mag\_auto\_z$<$2}, and the following
star-galaxy separation cut: \texttt{spread\_model\_i + (5./3.)
  spreaderr\_model\_i)$>$0.007}.  The flag cuts are DES-specific
  and described in \citet{DrlicaWagner2017}, while the other cuts are
  based on \textsc{SourceExtractor} \citep{Bertin1996} columns. We
further require the errors on the galaxy magnitudes to be less than
0.1. After applying the depth mask as well as the \redmapper mask, the
total number of galaxies in this sample is 11,263,383. A random
catalog that uniformly samples the galaxy catalog mask is generated
from the intersection of the $i>21.5$ depth mask and the \redmapper
mask. The final area used is $\sim$1,297\;deg$^{2}$.

\subsection{The Weak Lensing Shear and Photo-z Catalogs}

For the lensing measurements performed in this work, we use the two
DES Y1 shear catalogs: \textsc{MetaCalibration} \citep{Huff2017} and
\textsc{im3shape} \citep{Zuntz2013}. Both catalogs are tested and described in detail in
\citet{Zuntz2017}. The two catalogs were generated using
completely independent pipelines; performing the
measurements using both catalogs is therefore a powerful test of weak lensing
systematics, as shown in e.g. \citet{Troxel2017}. The
\textsc{MetaCalibration} catalog contains 34.8 million galaxies,
roughly $60\%$ more than the \textsc{im3shape} catalog due to the fact
that \textsc{MetaCalibration} uses the combined information of the
$r$, $i$ and $z$-band images while \textsc{im3shape} only uses
$r$-band images. We present our main results using the
\textsc{metacalibration} catalog but have checked that the
\textsc{im3shape} measurements show consistent results. 

For the weak lensing measurement in this work, redshift information is
needed for each source galaxy.  We use the photo-$z$ catalog described in 
\citet{Hoyle2017}, which is based on the 
template-based Bayesian Photometric Redshifts (BPZ) algorithm
\citep{Benitez2000}. Following McClintock et al. (in prep), we use both the 
mean of the PDF as well as a random draw from the full PDF for each galaxy 
when estimating the weak lensing mass profile. We describe the procedure in 
detail in \Sref{sec:lensing_profile}.

Both the shear calibration biases associated with the shear catalogs, and biases in the 
photo-$z$ catalog are well characterised in \citet{Zuntz2017} and \citet{Hoyle2017}. We 
do not account for these in the modeling since it mainly contributes to a scale-independent 
multiplicative factor at $\sim2\%$ and does not impact the inference of the splashback feature.  

\begin{table*}
\begin{center}
\caption{Selection criteria and sample sizes for the cluster samples used in this work. The mean 
$M_{\rm 200m}$ mass is derived via the mass-richness relation of 
\citet{Melchior2016}.}
\begin{tabular}{l|c|c|c|c|c|c|c}
Sample         & $z$ selection & $\lambda$ selection & \# of clusters & $\langle z \rangle$ & $\langle \lambda \rangle$ & $\langle R_{\lambda} \rangle$ ($h^{-1}$Mpc)& $\langle M_{\rm 200m} \rangle$ ($10^{14}$ M$_{\odot}$) \\ \hline
Fiducial               &$0.2<z<0.55$  & $20<\lambda<100$ &3684 & 0.41 & 31.6        & 0.79 & 2.5 \\ 
Low-z                 &   $0.2<z<0.4$    & $20<\lambda<100$ & 1588 &  0.32 & 32.2   & 0.79 & 2.5 \\  
Mid-z                 &   $0.4<z<0.55$    & $20<\lambda<100$ & 2096 & 0.48  &31.1   & 0.78 & 2.5 \\ 
High-z                 &   $0.55<z<0.7$    & $20<\lambda<100$ & 1518 & 0.61 & 30.3 & 0.78 & 2.4 \\  
Low-$\lambda$   &   $0.2<z<0.55$    & $20<\lambda<28$ & 1964 & 0.41 &23.3    & 0.75 &1.8 \\ 
High-$\lambda$   &   $0.2<z<0.55$    & $28<\lambda<100$ & 1720 & 0.40  &41.1 & 0.83 &3.3 \\ 
\end{tabular}
\label{tab:clust_sample}
\end{center}
\end{table*}

\section{Formalism}
\label{sec:model}

To model the 3D density profile around clusters, we use the
analytical model profile of \citetalias{Diemer2014}, which was found to be a good
description of dark matter halos in simulations across a wide range of mass,
redshift and accretion rate. The model includes two components:
``collapsed'' matter that has passed through at least one orbital
pericenter and is in orbit around the halo, and ``infalling'' material
that is falling towards the halo but has not experienced an orbital
pericenter. The profile of the collapsed matter is modelled by a
truncated Einasto profile \citep{Einasto1965}, while the infalling
material is modelled by a power law. The truncation of the Einasto
profile accounts for the splashback feature and is modelled using
the $f_{\rm trans}(r)$ term below.

The complete model for the 3D density, $\rho(r)$, is
\begin{eqnarray}
\rho(r) &=& \rho^{\rm coll}(r) + \rho^{\rm infall}(r), \label{eq:rho} \\
\rho^{\rm coll} &=& \rho^{\rm Ein}(r) f_{\rm trans}(r), \\
\rho^{\rm Ein} &=& \rho_s \exp \left(-\frac{2}{\alpha}\left[\left( \frac{r}{r_{s}}\right)^{\alpha} -1\right] \right), \\
f_{\rm trans}(r) &=& \left[1 + \left(\frac{r}{r_t} \right)^{\beta} \right]^{-\gamma/\beta}, \label{eq:ftrans} \\
\rho^{\rm infall} &=& \rho_0 \left(\frac{r}{r_0} \right)^{-s_e}. \label{eq:2halo}
\end{eqnarray}
Since $r_0$ is completely degenerate with $\rho_0$, we fix $r_0 = 1.5
h^{-1} {\rm Mpc}$ throughout.  Also, \Eref{eq:2halo} differs from the
formalism in \citetalias{Diemer2014} slightly in that we model the
mean-subtracted density profile, so there is no term corresponding to the 
mean density. Using the mean-subtracted profile allows us to model 
the average profiles of clusters at different redshifts more easily.

In practice, we measure the 2D projected profile instead of the 3D
profile, so it is useful to compute the projected density,
$\Sigma(R)$, which is related to the 3D density by
\begin{equation}
\Sigma(R) = \int_{-h_{\rm max}}^{h_{\rm max}} dh \, \rho(\sqrt{R^2 + h^2}),
\label{eq:Sigma_of_R}
\end{equation}
where $R$ is the projected distance to the halo center and $h_{\rm
  max}$ is the maximum scale of integration. We set $h_{\rm max} =$ 40
$h^{-1}$Mpc and test in \Aref{sec:hmax} that changing $h_{\rm max}$
does not significantly affect the inferred value of the splashbak
radius. However, as also shown in \Aref{sec:hmax}, changing $h_{\rm
  max}$ does impact the inferred slope of the density profile,
especially at large distances. Our choice of $h_{\rm max}=$ 40
$h^{-1}$Mpc is sufficiently large so that increasing $h_{\rm max} $ by
50\% only changes the large-scale slope by $<1\sigma$.

The model formulated above was intended for fitting the distribution
of mass around halos in simulations.  Following \citetalias{More2016}
and \citetalias{Baxter2017}, we nevertheless apply the model above to
the measured {\it galaxy} distribution (replacing $\Sigma(R)$ with
$\Sigma_{g}(R)$, the galaxy density profile) as well as the mass
profile.  In this approach, the unknown details of galaxy bias are
absorbed into the model parameters. In addition, by adopting this
  model we have assumed that the average profile of clusters with a
  range of mass, richness, redshift and miscentering parameters (see
  \Sref{subsec:mis_center}) can be described by one effective
  $\Sigma(R)$ or $\Sigma_{g}(R)$ profile for the whole sample.

From weak lensing shear measurements, we derive the differential mass
profile $\Delta\Sigma(R)$: 
\begin{eqnarray}
\label{eq:delta_sigma}
\Delta \Sigma(R) &=& \bar{\Sigma}(R) - \Sigma(R), 
\end{eqnarray}
where $\Sigma(R)$ is the projected surface mass density, and $\bar{\Sigma}(R)$ is 
the average of $\Sigma(R)$ within the circle of radius $R$, i.e.
\begin{eqnarray}
\label{eq:mean_sigma}
\bar{\Sigma}(R) = \frac{2 \pi \int_0^R dR'\, R' \Sigma(R')}{\pi R^2}.
\end{eqnarray}

Eqs.~\ref{eq:delta_sigma} and \ref{eq:mean_sigma} make it clear that
the lensing profile $\Delta \Sigma(R)$ depends on the density profile
of the cluster all the way down to $R=0$. This is problematic since
the lensing measurements on small scales may be affected by
systematics and the halo density profile may depart from the simple
Einasto model at small scales as a result of baryonic effects. It is
therefore convenient to introduce a new parameter, $\mu$, defined by
\begin{eqnarray}
\label{eq:mu}
\mu = \int_{0}^{R_{\rm min}} dR'  R' \Sigma(R'),
\end{eqnarray}
where $R_{\rm min}$ can be set to the minimum scale at which
$\Delta\Sigma$ is measured; we set $R_{\rm min} = 0.2 h^{-1} {\rm Mpc}$.
The expression for $\Delta \Sigma$ can then be written as
\begin{eqnarray}
\label{eq:DeltaSigma_model}
\Delta \Sigma(R) = 2\frac{\mu + \int_{R_{\rm min}}^R dR'  R' \Sigma(R') }{R^2} - \Sigma(R).
\end{eqnarray}
Treating $\mu$ as a free parameter effectively removes any sensitivity
of the $\Delta \Sigma$ profile to $\Sigma(R < R_{\rm min})$, where there
are no measurements.

Eqs.~\ref{eq:rho}---\ref{eq:2halo} and \ref{eq:DeltaSigma_model} use many free
parameters to fit functions which are very smooth as a function of
radius (see Figs.~\ref{fig:splashback_fiducial} and
\ref{fig:splashback_lensing}).  Consequently, there may be significant
degeneracies between the various parameters.  We emphasize, however,
that our intention here is not to extract robust constraints on the
model parameters themselves.  Instead, the goal of the model fitting
is mainly to smoothly interpolate between the data points to enable
the computation of the logarithmic derivatives of the 3D density
profile.

We describe below two additional pieces of modelling that we
incorporate into the above formalism in order to capture two important
observational complications: cluster miscentering and weak lensing
boost factor.

\subsection{Cluster Mis-centering}
\label{subsec:mis_center}

We model the effects of miscentering following the approach of
\cite{Melchior2016} and \cite{Baxter2017}.  The miscentered density profile,
$\Sigma$, can be related to the profile in the absence of
miscentering, $\Sigma_0$, via
\begin{eqnarray}
\Sigma = (1-f_{\rm mis})\Sigma_0 + f_{\rm mis} \Sigma_{\rm mis},
\label{eq:miscenter}
\end{eqnarray}
where $f_{\rm mis}$ is the fraction of clusters that are miscentered,
and $\Sigma_{\rm mis}$ is the density profile of the miscentered
clusters. For clusters that are miscentered by $R_{\rm mis}$ from the
true halo center, the corresponding azimuthally averaged density profile is
\citep{Yang2006, Johnston2007}
\begin{eqnarray}
&&\Sigma_{\rm mis} (R|R_{\rm mis}) = \nonumber\\
&&\int_0^{2 \pi} \frac{d\theta}{2\pi} \Sigma_0\left(\sqrt{R^2 + R_{\rm mis}^2  + 2RR_{\rm mis}\cos \theta}\right). \nonumber \\
\end{eqnarray}
The profile averaged across the distribution of $R_{\rm mis}$ values is then
\begin{eqnarray}
\Sigma_{\rm mis}(R) = \int dR_{\rm mis} P(R_{\rm mis}) \Sigma_{\rm mis}(R|R_{\rm mis}),
\end{eqnarray}
where $P(R_{\rm mis})$ is the probability that a cluster is
miscentered by a (comoving) distance $R_{\rm mis}$.  Following
\citet{Rykoff2016}, we assume that $P(R_{\rm mis})$ results from a
miscentering distribution that is a 2D Gaussian on the sky.  The 1D
$P(R_{\rm mis})$ is then given by a Rayleigh distribution:
\begin{eqnarray}
P(R_{\rm mis}) = \frac{R_{\rm mis}}{\sigma_R^2} \exp\left[ -\frac{R_{\rm mis}^2}{2\sigma_R^2} \right],
\label{eq:Pmis}
\end{eqnarray}
where $\sigma_R$ controls the width of the
distribution. \citet{Rykoff2016} assumed $\sigma_R = c_{\rm mis}
R_{\lambda}$, where $R_{\lambda} = (\lambda/100)^{0.2} \, h^{-1} {\rm
  Mpc}$, and used a combination of X-ray and SZ data to measure
$\ln(c_{\rm mis})=-1.13\pm 0.22$ and $f_{\rm mis} = 0.22\pm 0.11$. We
introduce $c_{\rm mis}$ and $f_{\rm mis}$ as free parameters in our
analysis of both the galaxy density and lensing profiles, imposing
priors corresponding to the \citet{Rykoff2016} constraints. We adopt
the mean value of $R_{\lambda}$ for our sample when computing
miscentering corrections, as listed in \Tref{tab:clust_sample}.

\subsection{Weak Lensing Boost Factor}
\label{sec:boostfactor}

Galaxies that are included in the shear catalog but are not behind the galaxy clusters of our 
sample will not be lensed and will therefore dilute the inferred shear. Since clusters contain 
many galaxies, the odds of such an occurrence increases towards the cluster center, resulting 
in systematic underestimation of the true $\Delta \Sigma$ profile. One typically calculates a 
{\it boost factor} to correct for this systematic \citep{Sheldon2004}.  
Our boost factor model is derived the same way as in McClintock et al. (in prep). We calculate 
the lensing-weighted average $p_{\rm phot}(z| R_{\rm clust})$ of source galaxies as a function 
of cluster centric radius and compare it with the corresponding reference $p_{\rm phot}(z| \rm{field})$ 
of field galaxies. The excess probability represents the member contamination in the source catalog. 
We then decompose the $p_{\rm phot}(z| R_{\rm clust})$ into two components: the reference distribution 
of field galaxies $p_{\rm phot}(z| \rm{field})$ and a Gaussian $p_{\rm phot}(z| \rm{Gauss})$ for the 
cluster member component. The decomposition is done jointly for all radial scales, such that consistency 
is enforced for the position and width of the gaussian $p_{\rm phot}(z| \rm{Gauss})$ components, and 
only the mixing amplitude at each scale, $A(R)$, is allowed to vary. $A(R)$ is then related to the 
traditional boost factor, $B(R)$, via $B(R) = 1/(1-A(R))$. The observed lensing signal, 
$\Delta\Sigma_{\rm  measured}(R)$, is then related to $B(R)$ via:
\begin{equation}
\Delta\Sigma_{\rm measured}(R) = \Delta\Sigma(R) / B(R).
\end{equation}  
At the minimum lensing scale considered in this work, $R = 0.2\, h^{-1}$Mpc, we find 
$B(R = 0.2\, h^{-1}{\rm Mpc})-1 \sim 0.26$; for scales greater than $1\,h^{-1}$Mpc, we find $B(R) - 1 < 0.05$.

\section{Measurement and Analysis}
\label{sec:measurements}

\subsection{Galaxy Density Profile, $\Sigma_g$}
\label{sec:gal_splashback}

We first measure the distribution of galaxies around the \redmapper
clusters. The galaxy distribution is expected to roughly trace
the matter distribution and has higher signal-to-noise than the weak
lensing measurements in our data. The density profile of galaxies
around a cluster is directly related to the galaxy-cluster correlation
function, $w(R)$, where $R$ is the projected comoving distance to the
cluster center. We work in {\it comoving} distances so that
$R=(1+z)R_{\rm phys}$, where $R_{\rm phys}$ is the projected physical
distance. 

As shown in \citetalias{Diemer2014}, after averaging over the
distribution of accretion rates at fixed halo mass, the location of
the splashback feature is expected to scale with physical
$R_{200m}$. Since the physical $R_{200m}$ is proportional to
$(1+z)^{-1}$ for fixed $M_{200m}$, measuring cluster-centric radii in
comoving units implicitly accounts for the redshift dependence of
physical $R_{200m}$ when stacking clusters of fixed $\lambda$ at
different redshifts. However, such scaling does not account for
potential systematic evolution of the mean halo accretion rate with
redshift.

Because our fiducial cluster sample includes a broad range of cluster
richnesses ($20<\lambda<100$), there will be some smearing of the
stacked signal due to variation in $r_{\rm sp}$ across the bin. In
\Aref{sec:lambda_scaling} we investigate the improvement in
signal-to-noise when we approximately scale each measurement in the
radial direction by the expected $R_{200m}$. Because the improvement
in signal-to-noise is modest and because scatter in the mass-richness
relation will complicate the relationship between the scaled
measurements and the simulations, our fiducial analysis does not
employ this scaling.

To measure the mean-subtracted galaxy density around the \redmapper
clusters, we divide the clusters into redshift bins of $\Delta z=0.05$
and measure the mean cluster-galaxy angular correlation function for
each bin $i$, $w(\theta, z_{i})$, using the Landy-Szalay estimator
\citep{Landy1993}. The angular correlation function $w(\theta, z_{i})$
is then converted into $w(R, z_{i})$, where $R$ is the projected
comoving distance for the angular separation $\theta$ at $z_{i}$, the
center of the redshift bin. The measurements for all the redshift bins
are combined by weighting $w(R, z_{i})$ with the number of cluster
random-galaxy random pairs $P_{\rm ran}^{i}$ in each bin. The
  $P_{\rm ran}^{i}$ values are calculated by counting the
  cluster-galaxy pairs in each angular bin using the cluster random
  catalog and the galaxy random catalog, then normalizing the pair
  counts by the number of clusters (galaxies) over the number of
  cluster randoms (galaxy randoms).
That is, we calculate
\begin{equation}
w(R) = \frac{\sum_{i} w(R, z_{i}) P_{\rm ran}^{i}} {\sum_{i} P_{\rm ran}^{i}}.
\end{equation}
Finally, to convert the measured correlation function into the
mean-subtracted density profile, $\Sigma_{g}(R)$, we multiply $w(R)$
by the mean density of galaxies around clusters,
$\bar{\Sigma}_{g}$. This is calculated via the weighted mean of the
mean galaxy density in each redshift bins $\Sigma_{g}^{i}$, where the
weight is the number of clusters $N_{c}^{i}$ in that bin. We have
\begin{equation}
\Sigma_{g}(R) = \bar{\Sigma}_{g}w(R),
\end{equation} 
where
\begin{equation}
\bar{\Sigma}_{g} = \frac{\sum_{i}\Sigma_{g}^{i} N_{c}^{i}}{\sum_{i} N_{c}^{i}}.
\end{equation} 

Since we only have photo-$z$ information for each galaxy, we cannot
select galaxies that are close to each cluster in redshift very
accurately. To avoid mixing galaxies with very different luminosities
across the full redshift range, we create an absolute-magnitude
limited sample following the approach of
\citetalias{More2016}. That is, before calculating the cluster-galaxy
cross-correlation in each redshift bin, we apply a luminosity 
cut on the galaxy sample where the absolute magnitudes are calculated
assuming all galaxies are at the same redshift as the clusters. In our
fiducial sample, this luminosity cut is $M^{*} \equiv
M-5\log(h)<-20.23$, where the upper limit is set to be the absolute
magnitude for galaxies with apparent magnitude $i=21.5$ at $z=0.55$.

The covariance between data points of different $R$ bins is derived
using 100 Jackknife samples, where the jackknife regions are derived
using the ``k-means'' method \citep{MacQueen1967}. 
The k-mean method splits the data points into groups, where 
the groups are divided so that the spatial coordinate of all the data points 
in each group is closest to the mean of them. With 100 jackknife
samples, each Jackknife region is approximately 3.7$\times$3.7
deg$^{2}$, which means we can reliably measure effects up to scales
$\sim$20 Mpc at the lowest redshift of interest $z=0.2$.

\subsection{Lensing Profile, $\Delta \Sigma$}
\label{sec:lensing_profile}

The tangential shear $\gamma_{t}$ of a background source galaxy around
a cluster is given by
\begin{equation} 
\gamma_{t} = -\gamma_{1} \cos(2\phi) -\gamma_{2} \sin(2\phi) \, ,
\label{eq:gammat}
\end{equation}
where $\phi$ is the position angle of the source galaxy with respect
to the horizontal axis of a Cartesian coordinate system centered on
the cluster and $\gamma_{1}$ and $\gamma_{2}$ are the two components
of shear measured with respect to the same coordinate system. For a
given lens redshift $z_{l}$ and source redshift $z_{s}$, the excess
surface mass density [see also \Eref{eq:delta_sigma}] is related to
the tangential shear according to
\begin{equation}
\Delta\Sigma (z_{l}, z_{s}; R) = \langle \gamma_{t}(z_{l}, z_{s}; R) \rangle \Sigma_{{\rm crit}} (z_{l}, z_{s}),
\label{eq:delta_sigma2}
\end{equation}
where $ \langle \gamma_{t}(z_{s}; R) \rangle$ is the mean tangential
shear for all lens-source pairs at these redshifts and $\Sigma_{{\rm
    crit}}$ is the critical surface density in comoving units defined
through
\begin{equation}
\Sigma_{\rm crit}^{-1} (z_l, z_s) = 4 \pi G (1+z_{l}) \chi(z_{l}) \left[ 1- \frac{\chi(z_{l})}{\chi(z_{s})}\right],
\end{equation}
where $G$ is the gravitational constant and $\chi(z)$ is the comoving distance to redshift $z$.
To combine the full source and lens redshift distributions, we follow the same approached used in 
McClintock et al. (in prep), where we measure 
\begin{equation}
\Delta \Sigma = \frac{\sum_{i,j} s_{i,j}\gamma_{i}^{t}}{\sum_{i,j}s_{i,j} \Sigma_{\rm crit}^{\prime -1}R_{\gamma,i}^{t}
+\left(\sum_{i,j}s_{i,j}\Sigma_{\rm crit}^{\prime -1} \right)\cdot \langle R_{s}^{t}\rangle}.
\end{equation}
The $i$ and $j$ subscripts denote the source galaxies and the lens galaxies, respectively. The weights 
for each source-lens pair, $s_{i,j}$, is calculated via
\begin{equation} 
s_{i,j} = \Sigma_{{\rm crit}}^{-1}(z_l, \langle z_s\rangle)_{i,j},
\label{eq:wj}
\end{equation}
where $z_{l}$ and $z_{s}$ are the mean redshift point-estimate provided by \textsc{BPZ}.
$\Sigma_{\rm crit}^{\prime}$ is the critical density calculated using a random redshift value 
drawn from the source $p(z)$. $R_{\gamma}$ and $R_{s}$ are the \textsc{MetaCalibration} responses to correct 
for the biased estimator and the selection bias on the ellipticity measurements.

For the covariance between data points in the different $R$ bins, we use the same 100 
Jackknife covariance as used in the galaxy measurements (\Sref{sec:gal_splashback}). 

\subsection{Model Fitting}
\label{sec:fitting}

We fit the models developed above to the data using a Bayesian
approach. We define a Gaussian likelihood, $\mathcal{L}$:
\begin{eqnarray}
\label{eq:likelihood}
\mathcal{L}(\vec{d} | \vec{\theta}) = \left[\vec{d} - \vec{m}(\vec{\theta}) \right]^T \mathbf{C}^{-1}  \left[ \vec{d} - \vec{m}(\vec{\theta})  \right],
\end{eqnarray}
where $\vec{d}$ is the data vector (either $\Sigma_{g}$ or $\Delta
\Sigma$), $\vec{m}(\vec{\theta})$ is the model vector (again either
for $\Sigma_{g}$ or $\Delta \Sigma$) evaluated at parameter values
$\vec{\theta}$, and $\mathbf{C}$ is the covariance matrix of the
data. The free parameters of the model are $\rho_0$, $\rho_s$, $r_t$,
$r_s$, $\alpha$, $\beta$, $\gamma$ and $s_e$, and the miscentering
parameters $f_{\rm mis}$ and $\ln c_{\rm mis}$. Additionally, when
fitting the $\Delta \Sigma$ data we also fit for $\mu$ as defined in
\Eref{eq:mu}.

Throughout this analysis we restrict the range of scales we use to fit the 
data to 0.1--10 $h^{-1}{\rm Mpc}$ for galaxies and 0.2--10 $h^{-1}{\rm
  Mpc}$ for lensing. We do not use the 0.1--0.2 $h^{-1}{\rm Mpc}$
scales for lensing in order to minimize sensitivity to the boost
factor.  Below $\sim 0.1 h^{-1}{\rm Mpc}$, the galaxy density
measurements are not well understood as the crowded nature of cluster
fields and the existence of the Brightest Cluster Galaxy (BCG) renders
various complications in the analysis on small scales related to
detection incompleteness, photometry inaccuracy and blending
\citep[][]{Melchior2015, Melchior2016}.  We restrict our fits to data
at scales less than $10 h^{-1}{\rm Mpc}$ since the model introduced in
\Sref{sec:model} breaks down at $\sim9R_{\rm vir}$, where the power
law of \Eref{eq:2halo} is no longer a good description of the
infalling term \citepalias{Diemer2014}. Since the mean virial radius
for the clusters in our sample is roughly 1 $h^{-1}$Mpc, we set the
upper limit to be $10 h^{-1}{\rm Mpc}$.

\begin{table}
\begin{center}
\caption{Priors used for the model fits of galaxy density and weak
  lensing mass profiles. The value in the parenthesis for
  $\log(\alpha)$ is only applied to lensing measurements. The ranges
  specified in brackets are for uniform priors while for the
  others we quote the mean and standard deviation of the Gaussian
  priors.}
\begin{tabular}{lc}
Parameter         & Priors   \\ \hline
$\rho_0$   &   $[0, 10]\,{\rm g/cm^3}$ \\
$\rho_s$  &   $[0, 10]\,{\rm g/cm^3}$ \\
$r_t$  &   $[0.1,5.0]\,h^{-1} {\rm Mpc}$\\
$r_s$   &  $[0.1,5.0]\,h^{-1} {\rm Mpc}$  \\
$\log(\alpha)$  & $\log(0.19)\pm$0.2 ($\log(0.19) \pm 0.1$) \\
$\log(\beta)$  &  $\log(6.0)\pm$0.2  \\
$\log(\gamma)$   & $\log(4.0)\pm$0.2    \\
$s_e$ & $[1,10]$ \\
$f_{\rm mis}$  &  0.22$\pm$0.11 \\
$\ln(c_{\rm mis})$  & -1.13$\pm$0.22 \\
\end{tabular}
\label{tab:priors}
\end{center}
\end{table}

We constrain the model parameters using a Markov Chain Monte Carlo (MCMC) algorithm. Following 
\citetalias{More2016} and \citetalias{Baxter2017}, we use the set of priors listed in \Tref{tab:priors}. For 
$\alpha$, we use a tighter prior than that used in \citetalias{More2016} and \citetalias{Baxter2017}, 
although these priors are still very wide compared to the $\alpha$ values seen in simulations \citep{Gao2008}. 
We have also checked that widening the $\alpha$ priors does not affect the resulting constraints on the 
splashback feature. For $\beta$ and $\gamma$, the priors are informative -- widening the priors leads to less 
constraining model fits. We discuss this point later in \Sref{sec:compare} but note that these 
priors still allow for a large range of profiles with and without the splashback feature. As such, the data (and not 
the prior) is still the main driver that determines the slope of the profile around the splashback radius. 
We also note there was a typographical error in \citetalias{More2016}. The 
actual Gaussian priors used were $\log(\beta) = \log(6.0) \pm 0.2$ and $\log(\gamma) = \log(4.0) \pm 0.2$, which 
is what is used in this work. For $f_{\rm mis}$ and $c_{\rm mis}$, we use values estimated in \citet{Melchior2016}. Otherwise, we implement 
the same priors as in \citetalias{More2016}. We sample the posterior of the parameters using the \texttt{emcee} 
code \citep{FM2013}. Convergence of the MCMC is assessed using trace plots and Geweke statistic.

For the remainder of the paper, we define the splashback radius
$r_{\rm sp}$ to be the radius at which the logarithmic derivative of
the 3D density profile $\rho(r)$ is at its minimum. To facilitate
comparison with previous literature, we also define $R_{\rm sp}$ to be
the location where the logarithmic derivative of the projected galaxy
density profile ($\Sigma_{g}$) has a minimum.

\section{Mass Profiles of \redmapper Clusters}
\label{sec:results}

In this section, we first present the measurements of the galaxy
density and lensing profiles around the fiducial cluster sample in
\Sref{sec:splashback_gal}.  In \Sref{sec:compare} we compare the galaxy
and lensing measurements, and discuss the implications. These
measurements are then compared to the measurements from dark matter
simulations in \Sref{sec:results_sims}. We follow by investigating the
redshift and richness dependencies of the splashback feature in
\Sref{sec:lamb_dep}.

\subsection{Galaxy and Lensing Profiles}
\label{sec:splashback_gal}

\begin{figure}
\centering
\includegraphics[width=0.95\linewidth]{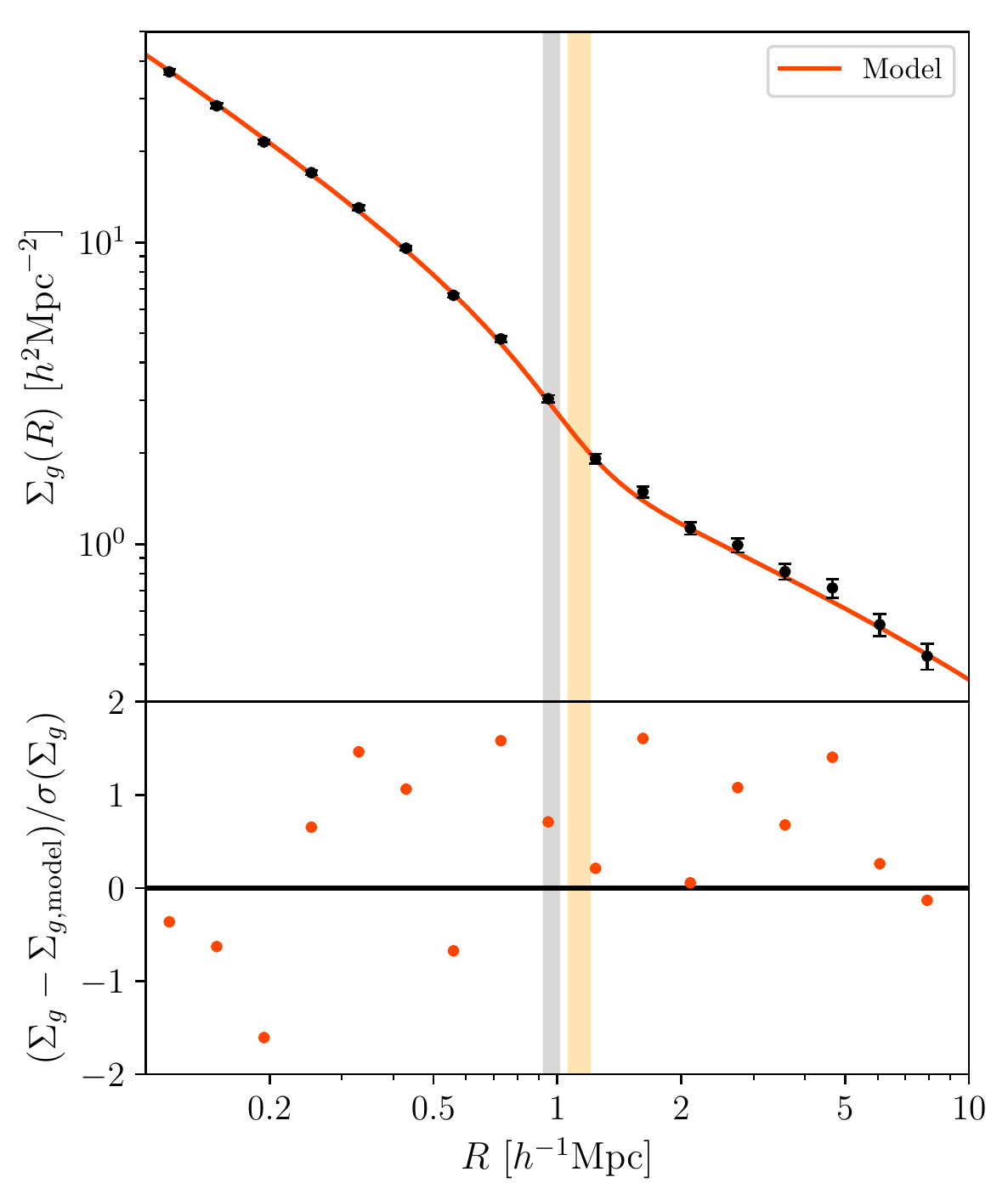}
\caption{\textit{Top:} The stacked surface density of DES Y1 galaxies
  around \redmapper clusters with $0.2 < z < 0.55$ and
  $20<\lambda<100$ (black points with error bars). The red line shows
  the model fit to the measurements. The inferred 3-dimensional
  $r_{\rm sp}$ is shown as the vertical orange band, with the width of
  the band indicating the $1\sigma$ uncertainty; the vertical grey
  band shows the inferred 2-dimensional $R_{\rm sp}$. We note that
  since the measurement is in projection, the grey band (instead of
  the orange band) indicates the point of steepest slope of the red
  line. \textit{Bottom:} the difference in the model and the
  measurements divided by the uncertainty in the measurement.  }
\label{fig:splashback_fiducial}
\end{figure}

The measurement of $\Sigma_g$ around \redmapper clusters in our
fiducial sample of $20<\lambda<100$ and $0.2<z<0.55$ is shown in the
top panel of \Fref{fig:splashback_fiducial}.  The red curve shows the
model fit of Eqs.~(\ref{eq:rho})--(\ref{eq:Sigma_of_R}) to the data
points with the inclusion of the miscentering prescription of
Eqs.~(\ref{eq:miscenter})--(\ref{eq:Pmis}).  The bottom panel of
\Fref{fig:splashback_fiducial} shows the residuals to the fit divided
by the uncertainty in the measurements. The residuals are consistent
with the uncertainties, indicating that the model is a good fit to the
data. The grey vertical band marks the steepest slope of the
$\Sigma_{g}$ model, which corresponds to the 2D splashback radius
$R_{\rm sp}$. The orange vertical band marks the steepest slope of the
inferred 3D density profile, $\rho(r)$, which corresponds to the
splashback radius $r_{\rm sp}$. Our measurement of the splashback
radius for the fiducial sample is consistent with that from
\citetalias{More2016} within 1$\sigma$ measurement uncertainty, which
provides a good confirmation of their results using an independent
data set and analysis pipeline\footnote{The redshift ranges of the
  clusters are somewhat different in the two analyses, but as we show
  in \Sref{sec:lamb_dep}, there does not appear to be significant
  redshift evolution in this range of redshifts.}.

\begin{figure}
\centering
\includegraphics[width=0.95\linewidth]{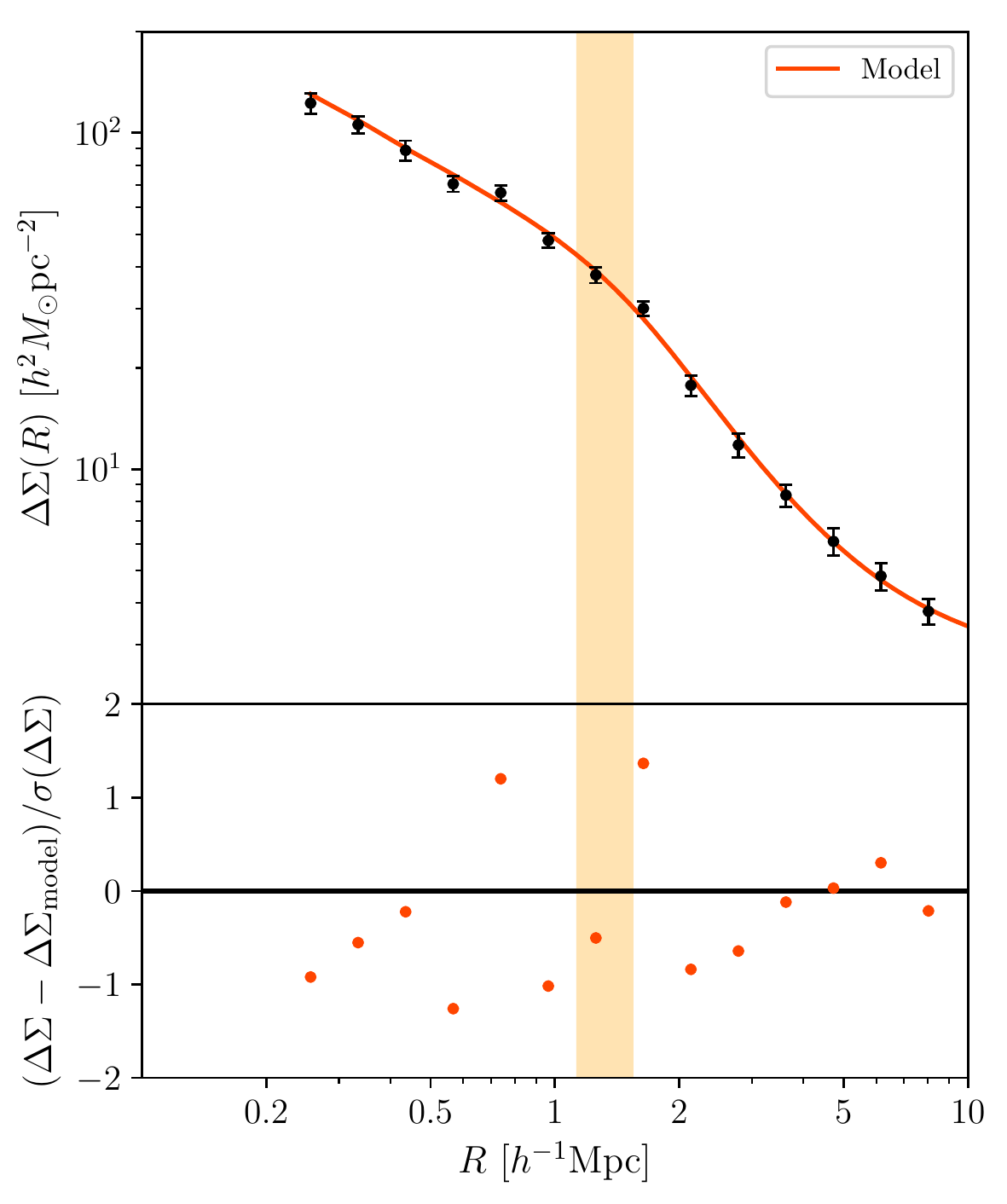}
\caption{\textit{Top:} The lensing stacked excess surface mass density
  around DES \redmapper clusters with $0.2 < z < 0.55$ and
  $20<\lambda<100$ (black points with error bars). The red line shows
  the model fit to the measurements.  The inferred 3-dimensional
  $r_{\rm sp}$ is shown as the vertical orange band, with the width of
  the band indicating the $1\sigma$ uncertainty.
\textit{Bottom:} the difference in the model and the
measurements divided by the uncertainty in the measurement.}
\label{fig:splashback_lensing}
\end{figure}

In \Fref{fig:splashback_lensing} we show the weak lensing measurement
of $\Delta \Sigma$ around our fiducial \redmapper clusters sample with
$20 <\lambda <100$ and $0.2<z<0.55$.  The top panel shows the data and
the model fit, while the bottom panel shows the residuals of the fit
divided by the uncertainty of the measurements. Again, the model
provides an excellent fit to the data. The corresponding $r_{\rm sp}$
is marked by the orange bands. Although the uncertainty in $r_{\rm
  sp}$ here is larger than that derived from the galaxy density
profile, we note that $r_{\rm sp}$ is very well constrained
\citep[compared to e.g.][]{Umetsu2017}.  The high signal-to-noise of
this measurement is a result of the combination of large number of
clusters and background source galaxies.

\subsection{Splashback Feature of \redmapper Clusters}
\label{sec:compare}

\begin{figure}
\centering
\includegraphics[width=0.99\linewidth]{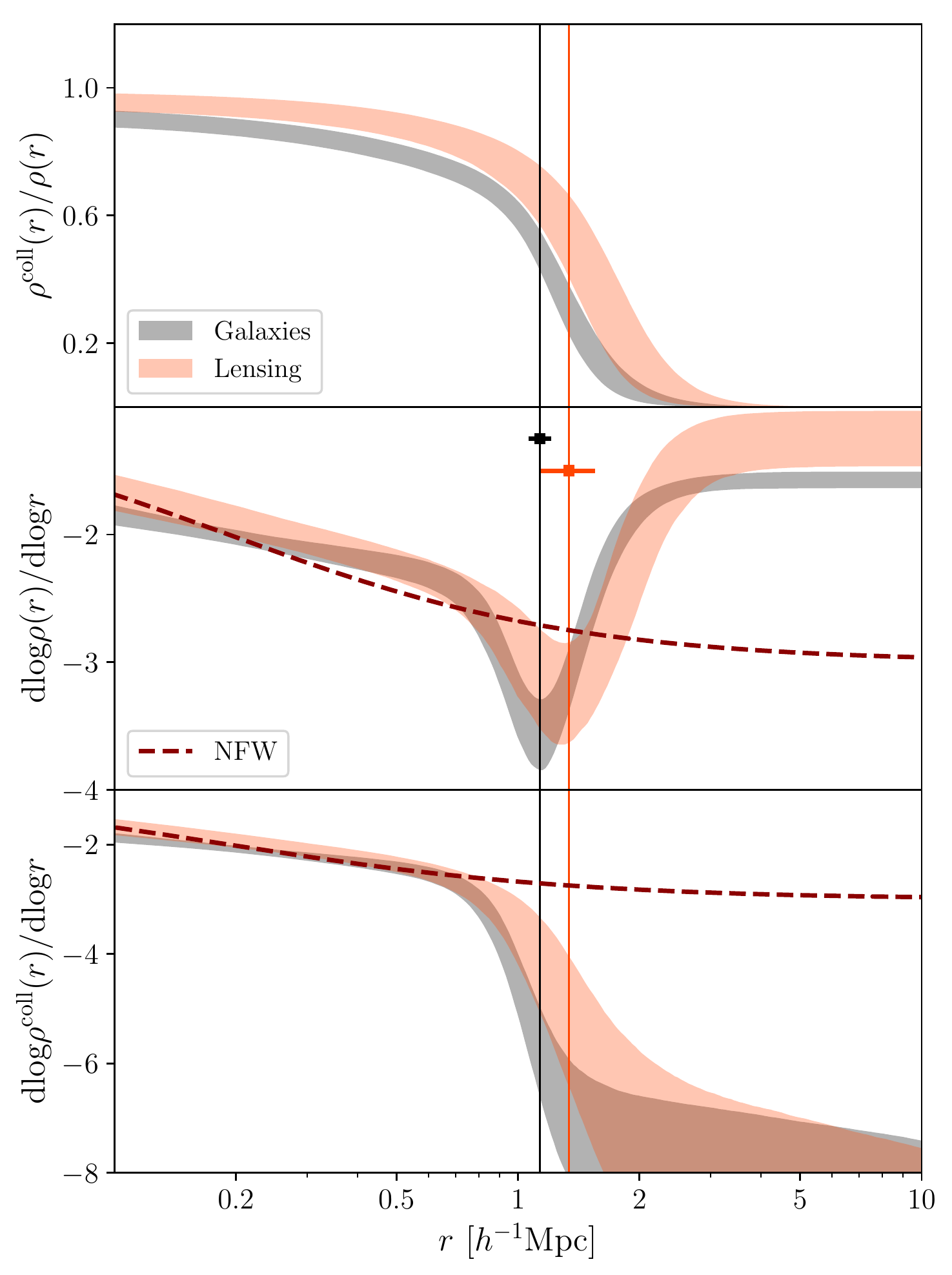}
\caption{Comparison of model-fit results from galaxy density
  $\Sigma_{g}$ (grey) and weak lensing $\Delta\Sigma$ (red).
  \textit{Top:} fraction of the density profile for the collapsed
  material over the total density profile. \textit{Middle:}
  logarithmic derivative of the total density profile compared to the
  logarithmic derivative of an NFW profile (dashed curve). \textit{Bottom:}
  logarithmic derivative of the profile for the collapsed material
  compared to the logarithmic derivative of an NFW profile.  The
  vertical lines mark the mean $r_{\rm sp}$ inferred from the model
  fits for both galaxy and lensing measurements, while the horizontal
  bars in the middle panel indicate the uncertainties on $r_{\rm sp}$.}
\label{fig:splashback_comparison}
\end{figure}

In Fig.~\ref{fig:splashback_comparison} we present the results of the
model fits to the galaxy density and weak lensing
measurements. Throughout the three panels, the vertical lines mark the
mean $r_{\rm sp}$ derived from the galaxy density (black) and the weak
lensing (red) profiles, whereas the horizontal bars in the middle
panel indicate the uncertainties (standard deviation of the $r_{\rm
  sp}$ distribution) of the two $r_{\rm sp}$ values.  The grey and red
bands show the 16 to 84 percentile confidence range for each
profile. The top panel shows the ratio $\rho^{\rm coll}(r)/\rho(r)$,
i.e. the fraction of the total density profile that is part of the
collapsed material profile. The difference in normalization between
the galaxy density and lensing measurements cancels in this ratio.  We
find that the galaxy density and lensing measurements yield very
consistent collapsed fractions, with the lensing measurements being
slightly higher. One might worry that we are drawing conclusions about
the collapsed material in a regime where it is completely dominated by
the infalling term. The top panel of \Fref{fig:splashback_comparison}
makes it clear that this is not the case: near $r_{\rm sp}$, the
collapsed profile term makes up 40--50\% of the total profile. Our
inferences about the collapsed profile in the transition regime is
therefore robust as long as the infalling material has a relatively
smooth profile, which is a good assumption here. Alternatively, one can introduce 
an additional ``limiting density'' term in the denominator as in Eq. 42 of \citet{Diemer2017b}, 
to avoid spurious contribution on small scales from the infalling term.

The middle panel of \Fref{fig:splashback_comparison} shows the logarithmic derivative 
of the total density profile inferred from the galaxy density and lensing measurements. 
The locations of the steepest slopes in the two profiles are consistent with the lensing 
measurement: the galaxy profile gives $r_{\rm sp}=1.13\pm 0.07$ $h^{-1}$Mpc and the 
weak lensing profile gives $r_{\rm sp}=1.34\pm0.21$ $h^{-1}$Mpc.
The amplitudes and 
shapes of the logarithmic derivative profiles are quite consistent, with the galaxy density 
profile slightly steeper at large radii. We find the total profile of both our galaxy and lensing 
measurements to be steeper than an NFW profile of similar mass at $r_{\rm sp}$ (as we 
discuss in more detail below). This is consistent with the expectation for a splashback feature.

An alternative is to look at the logarithmic slope of the collapsed profile, which is also 
the approach taken by \citetalias{Baxter2017}. This approach includes our model for the 
profile of the infalling material, which is assumed to be a power law. In the bottom panel of 
\Fref{fig:splashback_comparison} we show the logarithmic slope of the collapsed profile 
inferred from the galaxy density and lensing measurements. We find that at $r_{\rm  sp}$, 
the inferred collapsed profile from both galaxy and lensing profiles exhibit rapid steepening, 
achieving values much steeper than the slope of an NFW profile at scales around $r_{\rm sp}$ 
and beyond. This again is consistent with the picture that a splashback feature exists at 
the outskirts of these clusters.

\begin{figure}
\centering
\includegraphics[width=0.99\linewidth]{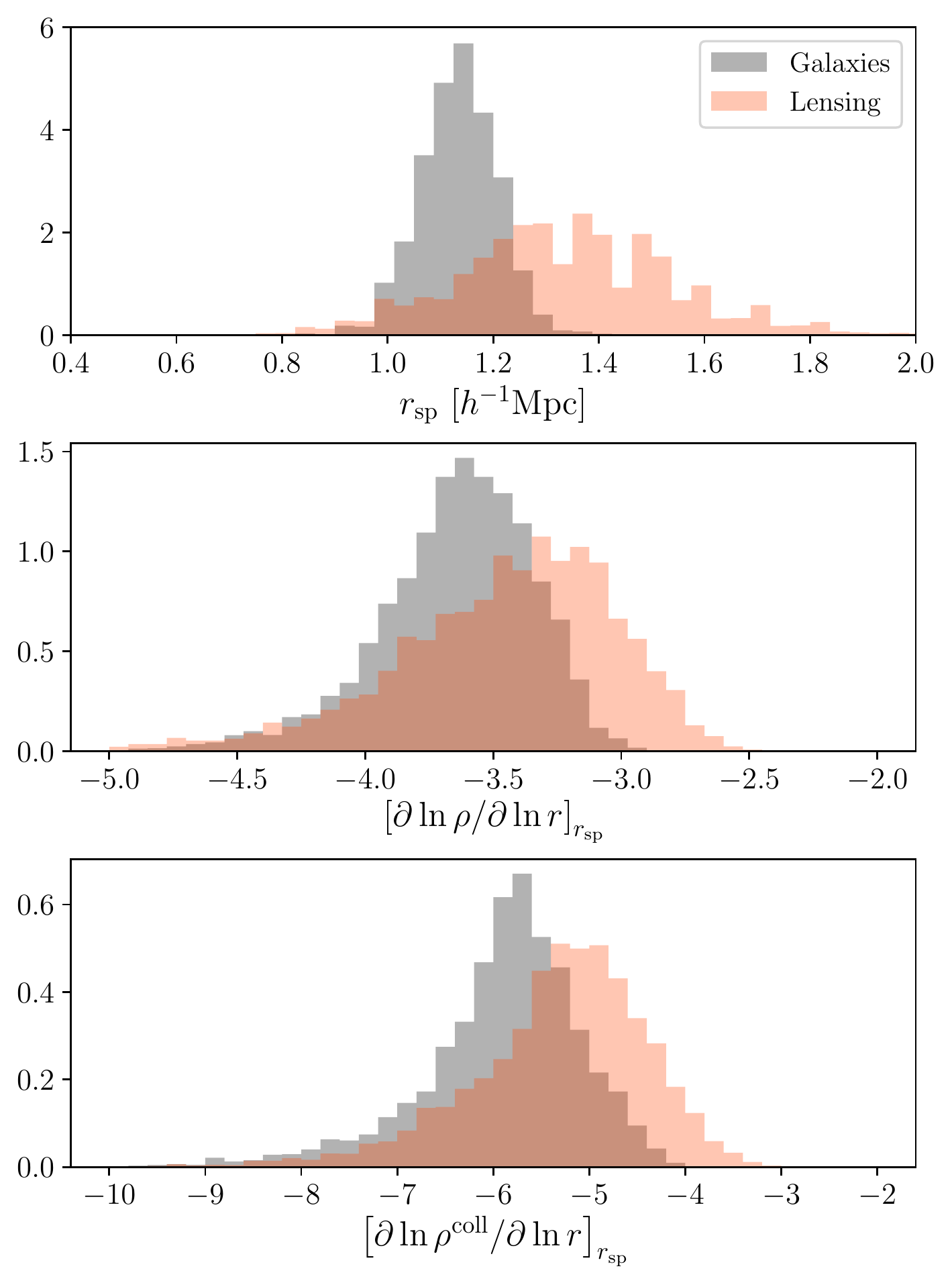}
\caption{\textit{Top:} posterior distributions of $r_{\rm sp}$ for
  galaxy density (grey) and lensing (red) data. \textit{Middle:}
  posterior distribution of the slope of the total matter profile at
  $r_{\rm sp}$. \textit{Bottom:} posterior distribution of the slope
  of the collapsed matter at $r_{\rm sp}$. All the distributions are
  marginalized over all nuisance parameters. }
\label{fig:rsp_distributions}
\end{figure}

The posterior distributions of $r_{\rm sp}$ and the slope of the total profile and the collapsed 
profile in \Fref{fig:splashback_comparison} are shown in \Fref{fig:rsp_distributions}. Here we 
clearly see that the galaxy and lensing measurements of $r_{\rm sp}$ and the slopes of the 
profiles are consistent with each other, with the lensing measurements having larger uncertainties. 
The measured logarithmic slope of the total profile at $r_{\rm sp}$ is $-3.6 \pm 0.3$ and 
$-3.5 \pm 0.4$ for the galaxy density and lensing profiles, respectively. The measured logarithmic 
slope of the collapsed profile is $-5.9 \pm 0.7$ and $-5.3 \pm 0.9$ for the galaxy density and
lensing profiles, respectively. These measured slopes can be compared to the expectation for 
an NFW profile. For the NFW profile predicted by the mass-richness relation of \citet{Melchior2016}, 
the logarithmic slope at $r_{\rm sp}$ is $\sim-2.7$, while the maximum possible slope is -3. The slope
of the total profile is therefore steeper than NFW at roughly $3.0\sigma$ for the galaxy density 
measurements, and $2.0\sigma$ for the lensing measurements. However, the NFW profile does not fully 
capture the contribution from infalling material near the cluster, which generically makes the profile less 
steep at $r_{\rm sp}$. Comparing the slope of {\it only the collapsed} component to that of the NFW 
profile, we find that it is steeper than NFW by $4.6\sigma$ for the galaxy density
profile and $2.9\sigma$ for the lensing profile. The values of $r_{\rm sp}$ derived from the MCMC, 
as well as the model parameters are listed in \Tref{tab:fit_parameters}.

\begin{figure}
\centering
\includegraphics[width=0.8\linewidth]{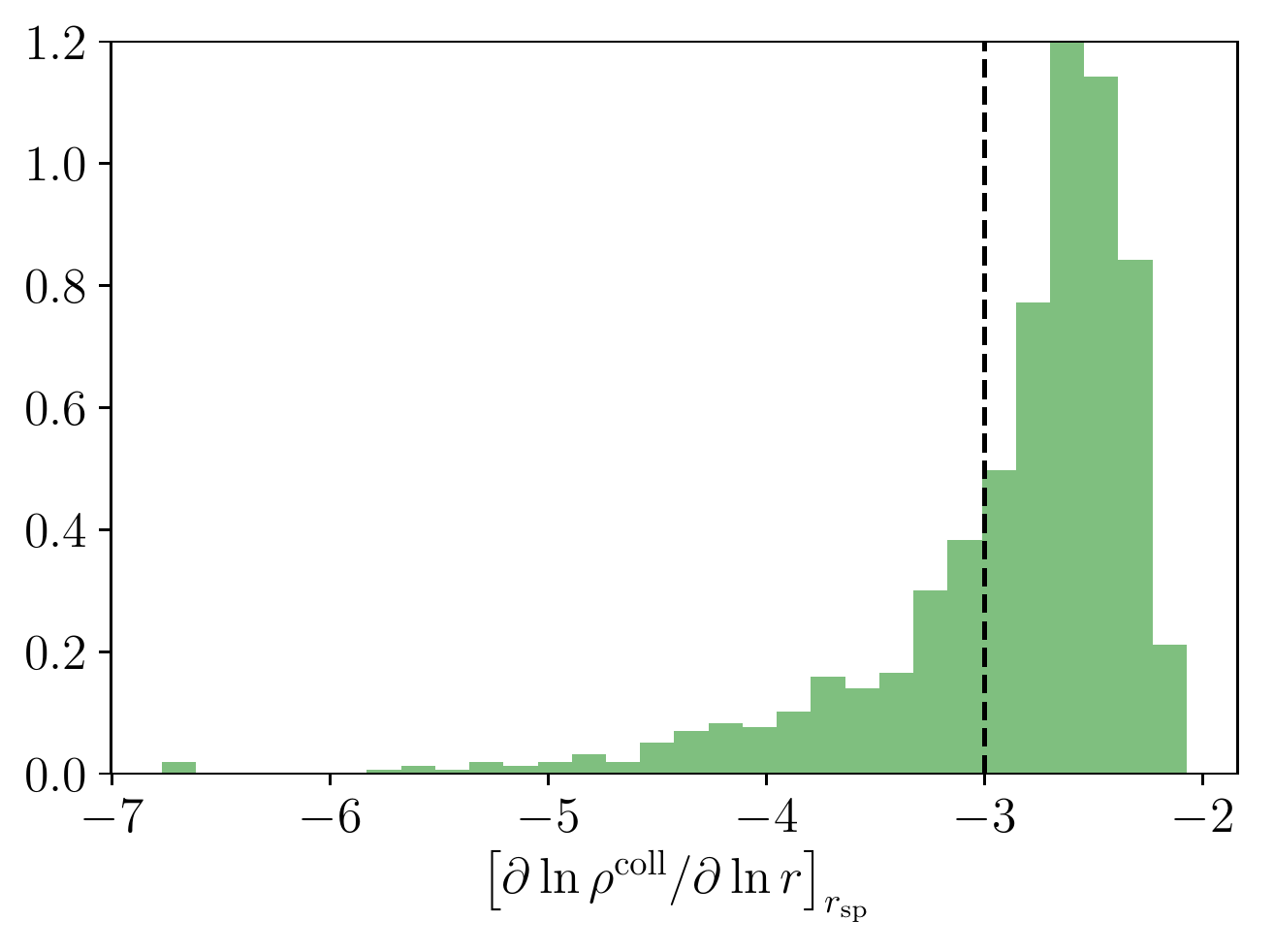}
\caption{Distribution of the logarithmic slope at $r_{\rm sp}$ when sampling over the prior distribution 
of several model parameters ($\alpha$, $\beta$, $\gamma$, $r_{t}$). The black dashed line indicates 
a rough indicator of the slope for an NFW profile. }
\label{fig:sample_priors}
\end{figure}

As discussed in \Sref{sec:fitting}, the parameters $\beta$ and
$\gamma$ are important for determining the behavior of the profile
around the splashback feature.  These parameters are degenerate, and
the priors that we place on them are informative.
To test how relaxing these priors would affect the splashback
measurement from lensing, we completely relax the $\gamma$ priors, and
examine the constraints on the slope of the profiles. We find the
slope of the total (collapsed) profile at $r_{\rm sp}$ to be
  $-3.7\pm0.6$ ($-6.2\pm2.0$) for the lensing measurement.  This
corresponds to a roughly 1.6$\sigma$ (1.8$\sigma$) steeper
profile compared to the NFW profile at $r_{\rm sp}$.  We also
  perform an additional check to see whether the priors are wide
  enough to span a range of profiles with and without a splashback
  feature ``detection''. That is, we check that the priors are not
  driving us to falsely detect a splashback-like steepening. To check
  this, we sample the priors of $\alpha$, $\beta$, $\gamma$, and $r_t$
  (the most relevant parameters for the splashback feature), generate
  model profiles and measure the slope of the profile at $r_{\rm
    sp}$. The resulting slope distribution is shown in
  \Fref{fig:sample_priors}. Noting that the minimum logarithmic slope
  achieved by an NFW profile is -3, we see that the priors allow
  profiles with slope both shallower and steeper than NFW.

\begin{table*}
\begin{center}
\caption{Model parameters for $\Sigma_{g}$ and $\Delta\Sigma$. $r_{s}$, $r_{t}$ and $r_{\rm sp}$ are 
in units of $h^{-1} {\rm Mpc}$. $r_{\rm 200m}$ is calculated using the mean mass and redshift for each 
cluster sample as listed in \Tref{tab:clust_sample}.}
\begin{tabular}{cccccccccc}
\hline
 & Sample & $\log(r_s)$ & $\log(r_t)$ & $\log(\alpha)$ & $\log(\beta)$ 
& $\log(\gamma)$ & $s_{\rm e}$ & $r_{\rm sp}$ & $r_{\rm sp}/r_{\rm 200m}$ \\ \hline
$\Sigma_{g}$ & Fiducial & $-0.82 \pm 0.10$ & $0.03 \pm 0.05$ & $-0.83 \pm 0.12$ & $0.92 \pm 0.14$ & $0.70 \pm 0.15$ & $1.57 \pm 0.07$  & $1.13 \pm 0.07$ & $0.82 \pm 0.05$ \\ 
 & Low-z & $-0.65 \pm 0.12$ & $-0.01 \pm 0.05$ & $-0.84 \pm 0.14$ & $0.92 \pm 0.17$ & $0.64 \pm 0.16$ & $1.61 \pm 0.09$  & $1.07 \pm 0.09$ & $0.73 \pm 0.06$ \\ 
 & Mid-z  & $-0.78 \pm 0.09$ & $0.07 \pm 0.09$ & $-0.72 \pm 0.14$ & $0.82 \pm 0.16$ & $0.66 \pm 0.17$ & $1.62 \pm 0.09$  & $1.12 \pm 0.14$ & $0.85 \pm 0.11$ \\ 
 & High-z  &$-0.66 \pm 0.12$ & $0.04 \pm 0.11$ & $-0.73 \pm 0.15$ & $0.86 \pm 0.18$ & $0.60 \pm 0.20$ & $1.46 \pm 0.11$  & $1.14 \pm 0.15$& $0.96 \pm 0.13$ \\ 
 & Low-$\lambda$ & $-0.75 \pm 0.09$ & $0.01 \pm 0.06$ & $-0.73 \pm 0.15$ & $0.91 \pm 0.16$ & $0.73 \pm 0.17$ & $1.50 \pm 0.08$  & $1.05 \pm 0.09$& $0.85 \pm 0.07$ \\ 
 & High-$\lambda$ & $-0.81 \pm 0.10$ & $0.10 \pm 0.08$ & $-0.83 \pm 0.12$ & $0.79 \pm 0.14$ & $0.62 \pm 0.17$ & $1.53 \pm 0.08$  & $1.27 \pm 0.14$ & $0.83 \pm 0.09$ \\ 
 $\Delta\Sigma$& Fiducial & $-0.62 \pm 0.15$ & $0.15 \pm 0.11$ & $-0.71 \pm 0.10$ & $0.75 \pm 0.17$ & $0.72 \pm 0.18$ & $1.23 \pm 0.24$  & $1.34 \pm 0.21$& $0.97 \pm 0.15$ \\  \hline

\end{tabular}
\label{tab:fit_parameters}
\end{center}
\end{table*}

\subsection{Comparison to N-body Simulations}
\label{sec:results_sims}

We now compare the galaxy density and lensing measurements around
\redmapper clusters to similar measurements made using dark
matter-only N-body simulations.  For this purpose we use the MultiDark
Planck 2 simulation from the CosmoSim database \citep[][also see
  \texttt{www.cosmosim.org}]{Prada2012, Riebe2013}.

Using the \textsc{Rockstar} \citep{Behroozi2013} halo catalogs made
available by CosmoSim, we identify a set of halos that is matched to
the \redmapper cluster catalog used in this work.  We match the
\redmapper clusters to the simulated dark matter halos on the basis of
halo mass and redshift. Using the best-fit mass richness relation from
\citet{Melchior2016}, we calculate the mean $M_{\rm 200m}$ halo mass of
our fiducial sample to be 2.5$\times10^{14}$ M$_{\odot}$. We then
determine in the simulations a mass threshold, $M_{\rm min}$, such
that the mean $M_{\rm 200m}$ mass of simulated halos between $M_{\rm min}$
and $10^{15}h^{-1}M_{\odot}$ is equal to $\bar{M}$.  We find $M_{\rm
  min}=1.0\times10^{14}$ M$_{\odot}$. The upper mass limit here has
little impact on our results, but ensures that a very small number of
extremely massive halos is not skewing our predictions. A total of
11,745 clusters are used in the simulations, and all are at a single
redshift $z=0.52$\footnote{The mean redshift of our cluster sample is
  $z=0.41$. We have chosen the closest redshift slice at $z=0.52$ in
  our simulations for an approximate comparison with the data. We
  however do not expect the subhalo profiles to vary significantly
  over this redshift range.}. We note that as we are employing the
mass-richness relation from a different sample (DES SV) and that there
is approximately a 10\% scatter on the mass-richness relation, we can
imagine that the masses in \Tref{tab:clust_sample} could be over/under
estimated. If the mass estimates were off by 1$\sigma$, the inferred
$r_{\rm sp}$ will move by $\sim3\%$.

The matching of our galaxy sample to objects in the MultiDark
simulations is more complicated since the simulations only contain
dark matter. We use both dark matter subhalos and dark matter
particles to perform the comparison with the galaxy and lensing
measurements. The connection between galaxies and subhalos depends on
a combination of environmental parameters \citep{Reddick2013,
  ChavesMontero2016}. Furthermore, subhalo density profiles around
massive halos are known to be flatter on small scales compared to
galaxy density profiles 
tidally stripped of mass near the cluster center (plus there can be
numerical artifacts on these scales in the simulations). Galaxies, on
the other hand, tend to live at the centers of their dark matter halos
and are therefore less likely to be tidally stripped
\citep[e.g.][]{Nagai:2005, Budzynski2012}.  These effects result in
differences between the matter, subhalo and galaxy profiles, but
mainly in the inner regions of the parent halo, well below the
splashback radius.

\begin{figure}
\centering
\includegraphics[width=0.99\linewidth]{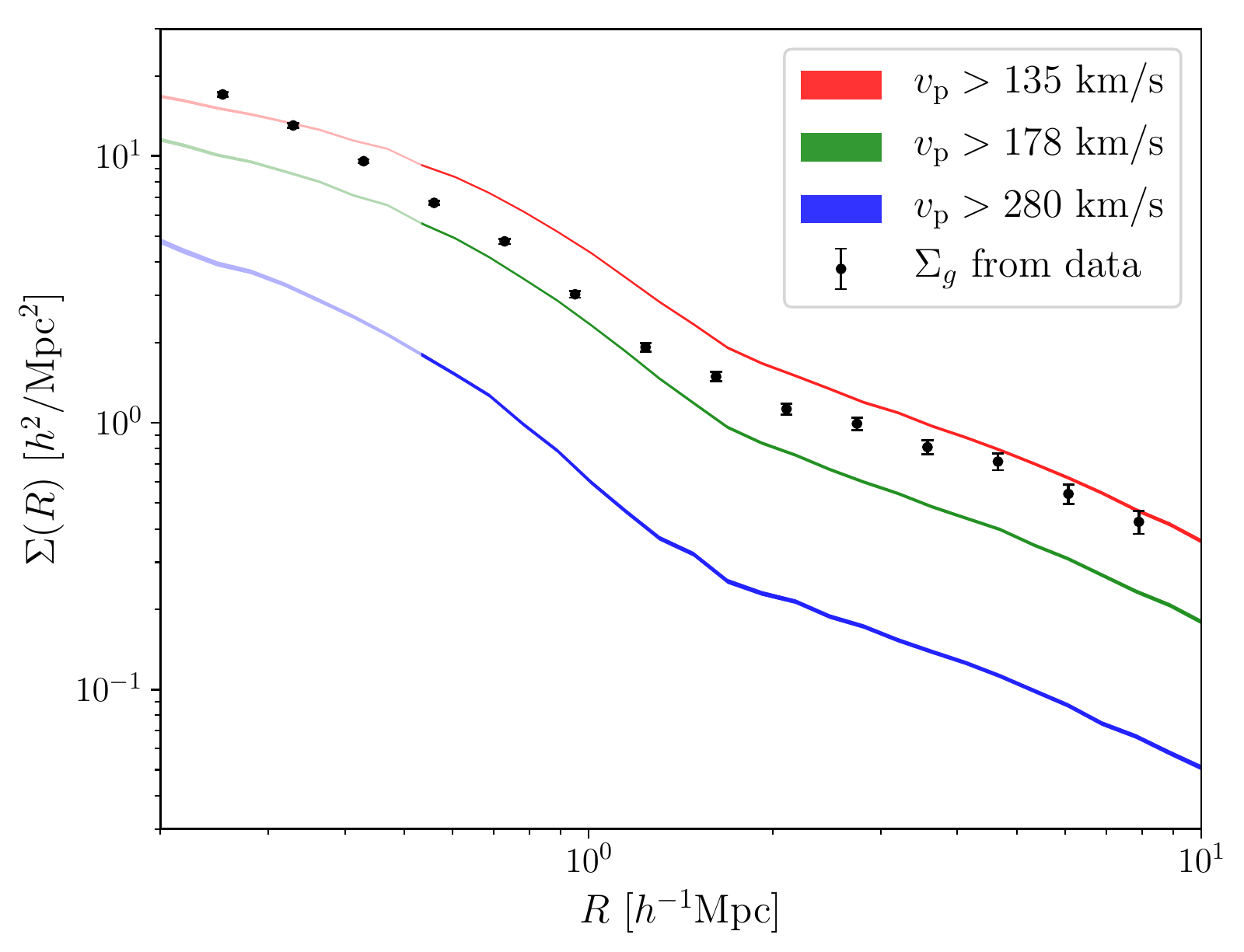}
\caption{Subhalo density profiles measured in simulations around halos with mass similar to that of our fiducial 
cluster sample. Different colors correspond to different choices of subhalo $v_P$. The data points are the 
galaxy profile measured with our fiducial sample, which lie between the two lower mass subhalo samples. 
The light shaded curves indicate the range excluded from the model fits described in \Sref{sec:results_sims}.}
\label{fig:splashback_sim_compare1}
\end{figure}

To construct a subhalo sample corresponding to our galaxy sample, we select Rockstar 
\citep{Behroozi2013}-identified subhalos using $v_{\rm p}$, the largest circular velocity attained by the 
subhalo over its history, which corresponds roughly to a mass cut on the subhalos at the time of accretion 
\citep{Reddick2013}. Unlike selection based on subhalo mass, selection on $v_{\rm p}$ is expected to lead 
to a sample that more closely approximates real galaxies since it is unaffected by tidal stripping of mass from 
the subhalo. The subhalo density profiles, $\Sigma_{\rm sub}$, for the simulated subhalos around dark matter 
halos of $M_{\rm min}<M<10^{15}h^{-1}M_{\odot}$ in the Multidark Simulations are shown in 
\Fref{fig:splashback_sim_compare1}, overlaid with the data $\Sigma_{g}$ measurements. At large scales, 
the amplitude of the $\Sigma_g$ curve scales with the abundance and can be compared with the galaxy 
density profile to find the approximate subhalo mass corresponding to a given galaxy sample. We consider 
three $v_{\rm p}^{\rm min}$ values: 135 km/s, 178 km/s, 280 km/s. In the simulations, there are on average 
28, 17, and 7 subhalos per cluster within 1.5 $h^{-1}$Mpc of the halo center for the three $v_{\rm p}$ cuts, 
respectively. As seen in \Fref{fig:splashback_sim_compare1}, 
the galaxy sample in our data lies between the two subhalo samples $v_{\rm p}^{\rm min}=135$ km/s and
$v_{\rm p}^{\rm min}=178$ km/s. 
That is, we can identify our galaxy sample with subhalos that are less massive then the $v_{\rm p}^{\rm min}=178$ 
km/s sample and more massive than the $v_{\rm p}^{\rm min}=135$ km/s sample. Note that we do not 
employ a rigorous abundance-matching procedure similar to \citetalias{More2016}. As a result, the amplitudes of 
our data points in \Fref{fig:splashback_sim_compare1} and later in \Fref{fig:dynamicalfriction_highlambda} do not 
match the subhalo profiles exactly.

\begin{figure}
\centering
\includegraphics[width=0.99\linewidth]{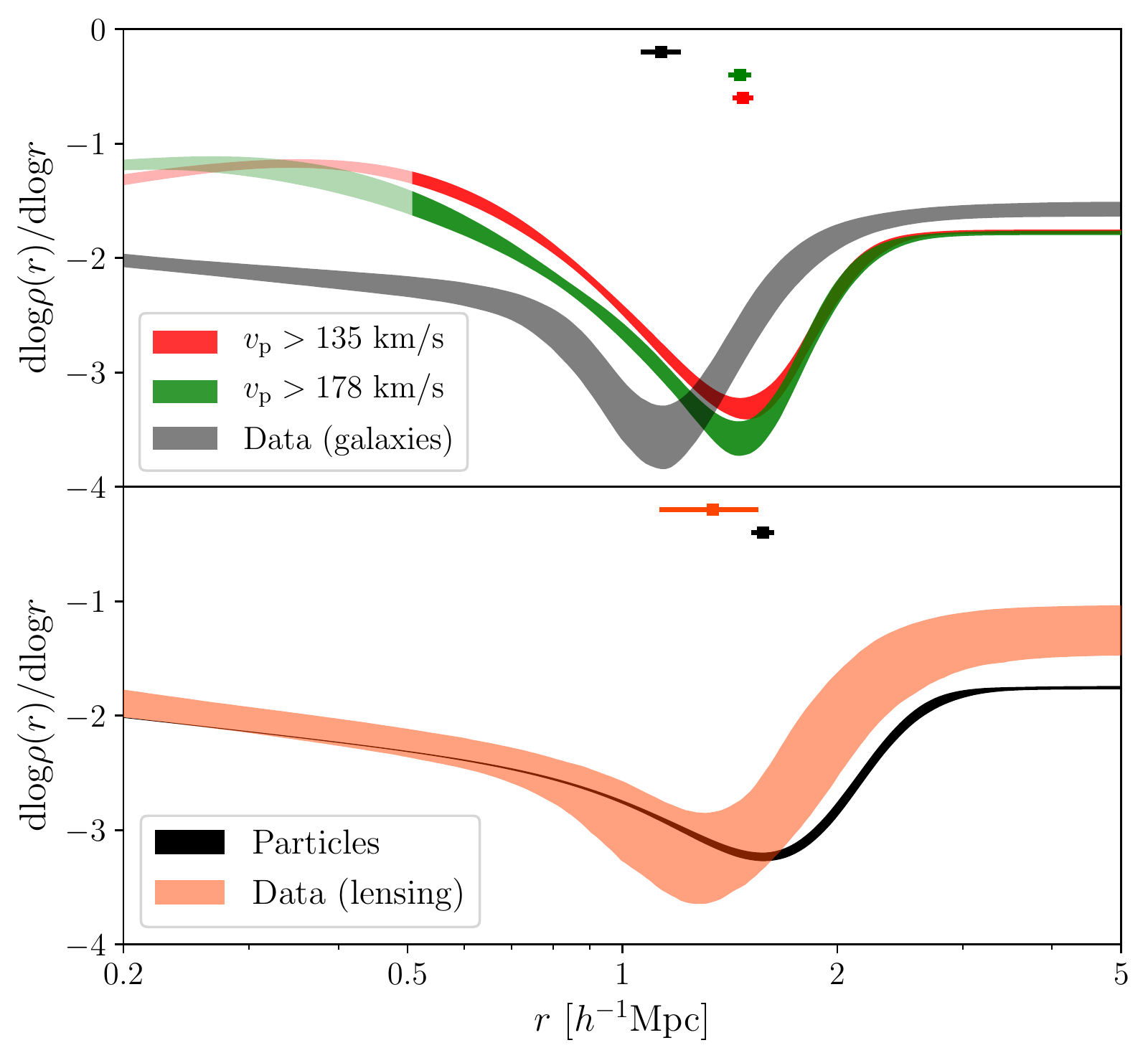}
\caption{Comparison of measurements from dark matter simulations and
data. \textit{Top:} the log-derivatives of the model fit to the
galaxy profiles in data and the subhalo profiles in simulations. The horizontal bars 
in each panel indicate the inferred location and uncertainty of $r_{\rm sp}$. Note that $r_{\rm sp}$ 
in the data is smaller than in the subhalo cases that are best matched to our galaxies. 
The faded section of the green and red curves indicate the regime where we expect differences between 
the data and simulations as we do not fit the subhalo profiles on small scales. 
\textit{Bottom:} same as top panel, but now comparing the slope of profile of the dark matter 
particles with the lensing measurements.}
\label{fig:splashback_sim_compare2}
\end{figure}

We fit the model described in \Sref{sec:model} to the subhalo profiles, excluding scales below 0.5 $h^{-1} 
{\rm Mpc}$ to minimize bias induced by the tidal stripping effect on small scales mentioned above. The model 
describes the subhalo profiles well after excluding the small scales. In the top panel of 
\Fref{fig:splashback_sim_compare2} we compare the logarithmic derivative of the model profile from our fiducial 
sample and from the two lower mass subhalo bins (since these bins bracket our galaxy sample). The inferred 
$r_{\rm sp}$ and uncertainty for each of the curves shown in the top panel of \Fref{fig:splashback_sim_compare2} 
are marked by horizontal bars on the top of the panel. As seen in the figure, the two lowest mass subhalo bins have 
essentially the same $r_{\rm sp}$, indicating that these subhalos are sufficiently small that they are not affected by 
dynamical friction. Since these two subhalo samples have masses that bracket that of our galaxy sample, we 
conclude that our measurements of $r_{\rm sp}$ from the galaxy density profile are not affected by dynamical friction. 
We will present a more thorough analysis of dynamical friction in \Sref{sec:df}.

The $r_{\rm sp}$ inferred from our galaxy density profile ($1.13\pm0.07$ $h^{-1}$Mpc) is significantly smaller 
than the corresponding subhalo measurements ($1.46\pm 0.05$ $h^{-1}$Mpc for the $v_{\rm p}^{\rm min}=178$ 
km/s subhalo sample), as seen in \Fref{fig:splashback_sim_compare2}. However, the steepest 
slope inferred from the simulations and data appear to be consistent, suggesting that we are seeing a level of 
steepening in the galaxy profile that is consistent with the splashback feature in simulations.
The overall shape of the galaxy profile in the data differs somewhat from that of subhalos in the simulations, where the 
small scale differences have been addressed above. These findings are consistent with those of \citetalias{More2016}.

On the bottom panel of \Fref{fig:splashback_sim_compare2}, we compare the lensing measurements with the 
dark matter particles. When fitting to the particle measurements we do not include the effects of miscentering. We 
find that the particles give consistent $r_{\rm sp}$ values as the two lower mass subhalo samples in the middle 
panel, and is larger than the lensing measurements by about 18\%. We note that the seemingly better 
agreement between the measurements and the simulations (about 1$\sigma$) is mainly driven by the fact that the 
lensing measurements have larger uncertainties. The slope of the lensing profile at large radii is shallower than the 
simulation particles; the same trend is seen in the galaxy vs. subhalo profiles. We have not investigated possible 
sources of this $\approx 2\sigma$ discrepancy. 

\begin{figure}
\includegraphics[width=0.92\linewidth]{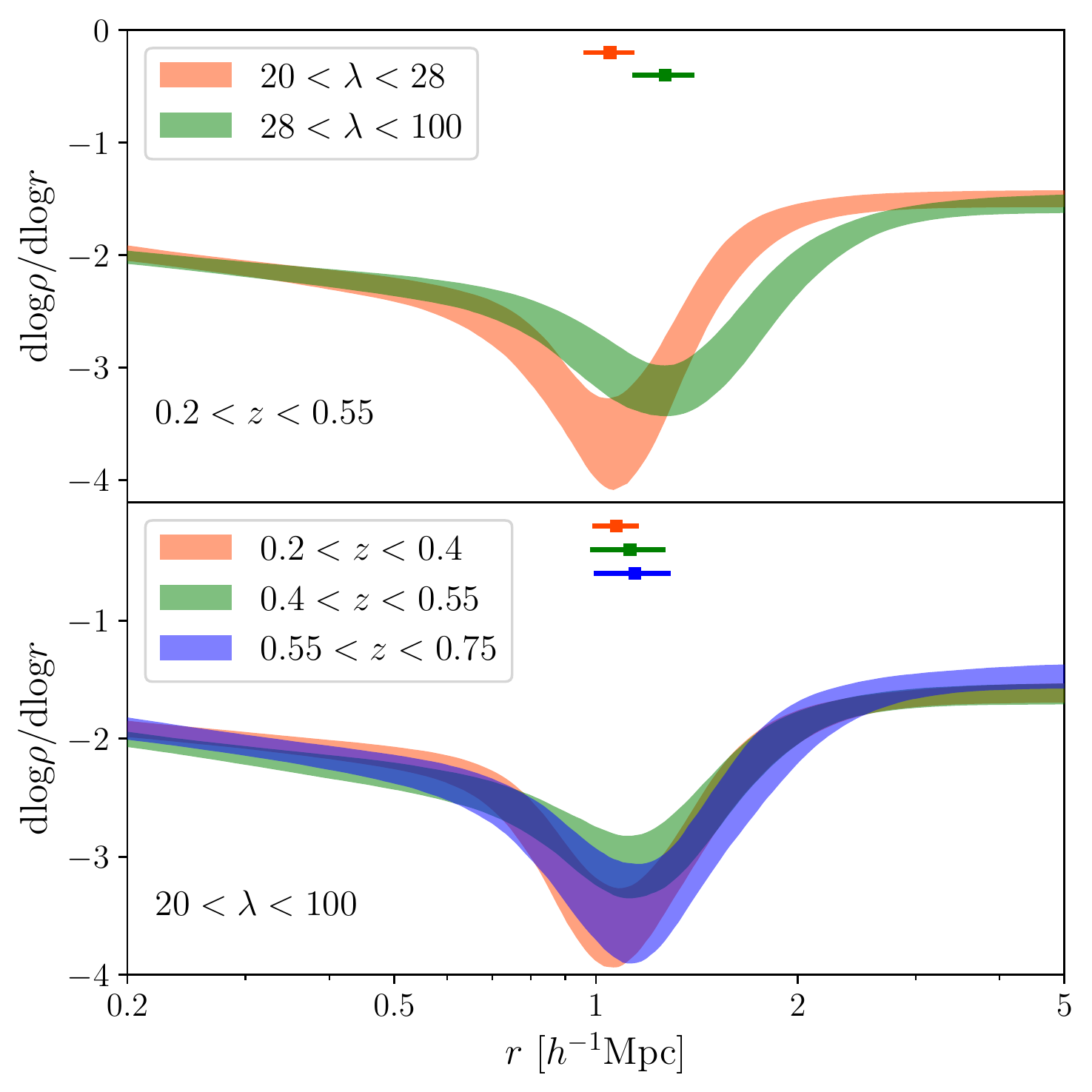}
\caption{\textit{Top:} logarithmic derivative of the model fits to the 
$\Sigma_{g}$ measurements with different richnesses. \textit{Bottom:} 
similar to the top panel but for different redshift bins. The horizontal bars 
in each panel indicate the inferred location and uncertainty of 
$r_{\rm sp}$ in the different subsamples.}
  \label{fig: splashback_zdep_ldep}
\end{figure}

\subsection{Richness and Redshift Dependences of $r_{\rm sp}$}
\label{sec:lamb_dep}

We now consider the richness dependence of the splashback feature. According to simulation tests in 
\citetalias{Diemer2014} and \citetalias{Adhikari2014}, one would expect the splashback feature to be 
shallower and appear at smaller scales for lower mass (or richness) clusters. We measure the richness 
dependence of the splashback location by dividing the fiducial cluster sample into 2 richness subsamples 
-- $20<\lambda<28$ and $28<\lambda<100$. The bins are chosen so that the number of clusters are 
approximately equal in both bins. The mean richness in the two bins are 23.3 and 41.1, respectively. In the 
top panel of \Fref{fig: splashback_zdep_ldep} we show the log-derivatives of the model fits to the galaxy 
density profiles of these two subsamples. We find that $r_{\rm sp}$ is 1.05$\pm$0.09 $h^{-1} {\rm Mpc}$ and 
1.27$\pm$0.14 $h^{-1} {\rm Mpc}$ for the low and high richness samples, respectively. The dependence of the 
mean $r_{\rm sp}$ on the mean $\lambda$ is roughly $r_{\rm sp} \propto \lambda^{0.33\pm0.24}$, which is 
consistent with expectation\footnote{Since $r_{\rm sp}\propto R_{\rm 200m}$, we expect $r_{\rm sp} \propto M_{\rm 200m}^{1/3}$. 
\citet{Melchior2016} found $M_{\rm 200m}\propto \lambda^{1.12}$, suggesting $r_{\rm sp} \propto \lambda^{0.37}$.}
from the slope of the mass-richness relation of \redmapper clusters measured in \citet{Melchior2016}, 
$r_{\rm sp} \propto \lambda^{0.37}$. We note, however, that detailed shapes of the logarithmic derivatives 
measured from the data exhibit some puzzling differences from simulations. In particular, we find that the 
high-richness cluster sample has a shallower splashback feature than the low-richness cluster sample. In 
the simulations of \citet{Diemer2014}, on the other hand, higher mass halos tend to have {\it sharper} splashback 
features.

In principle, our measurement of the richness dependence of the
splashback radius could be impacted by dynamical friction. As
discussed in \Sref{sec:intro}, dynamical friction will result in a
decrease in the observed splashback radius measured via the galaxy
density profile. This effect is expected to be weaker for larger host
halos, which could result in an increase in the observed scaling of
the splashback radius with mass relative to the expectation from
particles in simulations (the particle profile is not impacted by
dynamical friction). However, as we show in \S\ref{sec:df}, for our
fiducial galaxy sample, dynamical friction does not appear to have a
significant impact on the inferred splashback radius. Consequently,
our measurement of the richness dependence of the splashback radius
can be compared directly to the expectation from particles in
simulations.
      
We next consider the redshift dependence of the splashback feature. \citetalias{Adhikari2014} 
looked at the redshift dependence of the splashback feature in simulations, finding that for a 
given accretion rate, $r_{\rm sp}$ becomes larger at higher redshift, which results from a 
simple scaling with the background cosmology (specifically $\Omega_{m}$). When averaged 
over a distribution of accretion rates, however, \citetalias{Diemer2014} finds that the results are 
consistent with no redshift evolution. We test this by performing the same $\Sigma_{g}$ 
measurement in three redshift bins: $0.2<z<0.4$, $0.4<z<0.55$ and $0.55<z<0.75$. The lowest 
redshift bin is similar to that used in \citetalias{More2016}, whereas the highest redshift sample 
is not strictly volume-limited. In the bottom panel of \Fref{fig: splashback_zdep_ldep} we show the 
log-derivative of the model fit to the measurements for the three redshift bins, with the inferred 
$r_{\rm sp}$ marked on the plot and listed in \Tref{tab:fit_parameters}. We find no evidence of 
redshift evolution of $r_{\rm sp}$ over this redshift range. Given that we do not select the clusters 
in accretion rate, our finding of no redshift evolution is consistent with that found in \citetalias{Diemer2014}. 
One might worry 
that the mass-richness relation also evolves with redshift, which could complicate the comparison. 
However, in our sample we do not find a significant evolution of mass over the three redshift samples 
(see \Tref{tab:clust_sample}), which means we indeed do not see a redshift evolution of $r_{\rm sp}$ 
for fixed halo mass.

\section{Effect of Dynamical Friction}
\label{sec:df}

\begin{figure}
\center
\includegraphics[width=0.99\linewidth]{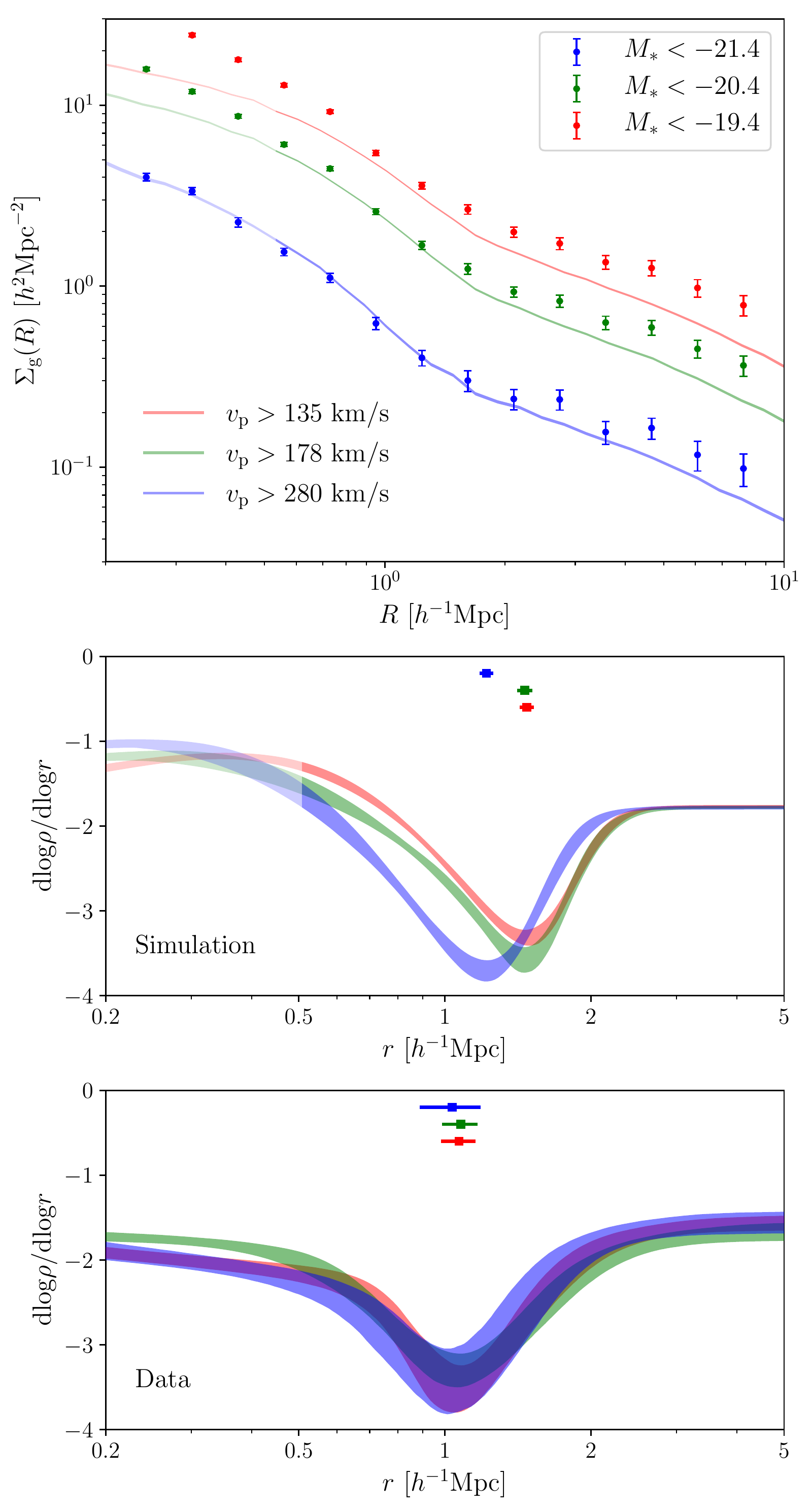}
\caption{Effects of changing the galaxy luminosity cut on the inferred $r_{\rm sp}$ around 
$20 < \lambda < 100$, $0.2<z<0.4$ clusters. \textit{Top:} measurement of $\Sigma_{g}$ 
profiles for the three luminosity bins (data points) and the three subhalo profiles in 
\Fref{fig:splashback_sim_compare1} (solid lines). The light shaded curves indicate the range 
excluded from the model fits. The subhalo samples are not abundance-matched to the galaxies, 
therefore we do not expect the amplitudes of the data points to agree with the solid lines. 
\textit{Middle:} log-derivatives of model fits to the subhalo density 
profiles measured in simulations from the top panel. The faded section of the curves indicate the 
regime where we expect differences between the data and simulations as we do not fit the subhalo 
profiles on small scales. \textit{Bottom:} the log-derivative of the model fits to the three galaxy density 
profiles. The horizontal bars indicate the inferred location and uncertainty of the 3-dimensional 
$r_{\rm sp}$ for each galaxy sample.}
\label{fig:dynamicalfriction_highlambda}
\end{figure}

As discussed in \Sref{sec:intro}, measuring the splashback radius provides an avenue 
for detecting the effects of dynamical friction in galaxy clusters. The rate of deceleration 
due to dynamical friction for a subhalo travelling through a cluster is proportional to the 
mass of the subhalo. Consequently, more massive (brighter) galaxies are expected to 
splashback at smaller radii. We first test this expectation in simulations by looking at the 
log-derivative of the model fits to the three $\Sigma_{\rm sub}$ curves in the upper panel
of \Fref{fig:splashback_sim_compare1}. The corresponding log-derivative profiles are 
shown in the middle panel of \Fref{fig:dynamicalfriction_highlambda}, together with the 
inferred $r_{\rm sp}$. It is clear that subhalos with $v_{\rm p} > 280\,{\rm km}/{\rm s}$ 
have a significantly smaller splashback radius than lower mass subhalos, the expected 
consequence of dynamical friction.  For $v_{\rm p} > 280\,{\rm km}/{\rm s}$ 
subhalos, we find $r_{\rm sp}=1.21$ $h^{-1}$Mpc, while for the other two subhalo 
samples we find $r_{\rm sp}=1.47$ $h^{-1}$Mpc -- this $\sim20\%$ difference is consistent 
with that found in \citet{Diemer2017}.

Since we cannot directly measure the masses of the galaxies in our sample, we divide the 
galaxies based on luminosity, which correlates with mass. We define three luminosity bins from 
our galaxy sample ($M_{*}<-19.4$, $M_{*}<-20.4$ and $M_{*}<-21.4$) and measure the resultant 
$\Sigma_{g}$ profiles around the low-z cluster sample ($20<\lambda<100$, $0.2<z<0.4$) as 
shown in the top panel of \Fref{fig:dynamicalfriction_highlambda}. We use the low-z sample so
that we can lower the luminosity cut on the galaxies and have higher signal-to-noise measurements. 
Overlaying the same subhalo profiles from the dark matter simulation as in 
\Fref{fig:splashback_sim_compare1}, we find that the brightest galaxy bin ($M_{*}<-21.4$) roughly 
corresponds to the most massive subhalo bin ($v_{\rm p}>280$ km/s), which is also the sample that 
has showed signs of dynamical friction. The two fainter galaxy bins roughly correspond to the two 
lower mass subhalo samples. We fit all three galaxy measurements to the same model used in
\Sref{sec:splashback_gal} and show the log-derivative profile of the models in the bottom panel of 
\Fref{fig:dynamicalfriction_highlambda}. The galaxies show similar behaviors to what was 
observed with the subhalos in the dark matter simulations -- the two fainter galaxy bins have 
consistent $r_{\rm sp}$ measurements, while the brightest galaxy bin has a slightly smaller 
$r_{\rm sp}$. However, the difference between the brightest galaxy bin and the other two bins is 
smaller than what is expected from the simulations and well within the measurement uncertainties.
Furthermore, as we show in \Aref{sec:rc_test}, this 
measurement is sensitive to the choice of the \redmapper parameter $R_{0}$ (see \Sref{sec:model_sys}).
This test does, on the other hand, confirm that our fiducial galaxy sample used in \Sref{sec:results} is not 
affected by dynamical friction. 

Comparing in more detail the bottom two panels of \Fref{fig:dynamicalfriction_highlambda}, we also 
find other qualitative differences in the profiles: the most massive galaxy sample shows a shallower 
log-derivative compared to the other two galaxy bins, which is in the opposite direction of what is 
expected from the subhalo simulations. To further investigate these subtle differences and systematics 
effects would require more realistic simulations that capture the baryonic physics on small scales. We 
defer this study to future work.

One can imagine further increasing the effect of the dynamical friction by going to lower-mass clusters, 
an approach taken by \citet{Adhikari2016}. This is because one expects the effect of dynamical friction 
to be larger for smaller host halos (the effect of dynamical friction scales with $M_{\rm sub}/M_{\rm host}$, 
where $M_{\rm sub}$ is the subhalo mass and $M_{\rm host}$ is the host halo mass). However, we note 
that the mass estimates for \redmapper clusters below $\lambda=20$ are less reliable as shown in 
\citet{Melchior2016}. We therefore do not perform further measurements using the low richness clusters. 
We also tested that our conclusion of this analysis does not change when using non-overlapping 
magnitude bins, which could in principle enhance the effect of dynamical friction.

\section{Potential biases due to the \redmapper algorithm}
\label{sec:model_sys}

One potential concern for splashback measurements relying on
\redmapper clusters was pointed out in recent work by
\citet{Busch2017}. \redmapper identifies clusters based on
overdensities of red galaxies on the sky. Selection effects in
\redmapper could therefore result in changes to the measured galaxy
density profile relative to the true galaxy density profile around
\redmapper clusters, which could potentially result in biases to
splashback measurements that use the galaxy density profile around
these clusters.  We review and investigate this issue below.

In the \citetalias{More2016} measurement of splashback with SDSS,
\redmapper-identified clusters were split into two subsamples based on
$\langle R_{\rm mem} \rangle$, defined as the average cluster-centric
distance of the cluster members weighted by membership
probability. These two samples were found to have significantly
different $r_{\rm sp}$ as well as large-scale clustering
amplitudes. \citet{Zu2016} and \citet{Busch2017} later pointed out
that $\langle R_{\rm mem} \rangle$ can be strongly affected by
projection effects due to the way \redmapper assigns members to
clusters. That is, clusters that live in dense environments are
likely to have a large number of spurious members from line-of-sight
projections that have low membership probability, but which contribute
to large $\langle R_{\rm mem} \rangle$ values. These clusters will
have a higher large-scale clustering amplitude as a result of their
association with projected structures along the line of sight.
Selecting on $\langle R_{\rm mem} \rangle$ can therefore result in
spurious assembly bias signals \citep{Miyatake2016,Zu2016}. Given the
sensitivity of $\langle R_{\rm mem} \rangle$ to projection effects, we
have not employed this quantity in our analysis.

Using the Millennium simulation \citep{Springel2005},
\citet{Busch2017} also argued that the aperture radius, $R_c$, used by
\redmapper to define cluster richness could impact the cluster density
profile and therefore the splashback feature. \redmapper computes the
cluster richness as a weighted sum over galaxies within $R_c$ of the
assumed cluster center, where
\begin{eqnarray}
R_c(\lambda) = R_0 \left( \lambda / \lambda_0\right)^{\beta}.
\label{eq:rc}
\end{eqnarray}
Values of $R_0 = 1.0\,h^{-1}{\rm Mpc}$, $\beta = 0.2$ and
$\lambda_0=100$ were chosen to minimize scatter in the mass-richness
relation \citep{Rozo2009}. \citet{Busch2017} considered the effects of
changing $R_0$ on the results of their simulated \redmapper
measurement\footnote{Since increasing $R_0$ necessarily means that
  clusters will have larger richness, \citet{Busch2017} simultaneously
  varied $R_0$ and $\lambda_0$.}. For catalogs generated with $R_0 =
0.67,1,1.5$ $h^{-1}{\rm Mpc}$, \citet{Busch2017} found that the
inferred $r_{\rm sp}$ and profile shapes were altered. Note that as
mentioned above, these analyses were carried out using the Millennium
simulation \citep{Springel2005} and a simplified procedure that
approximates the \redmapper algorithm. Consequently, the exact
quantitative effect on $r_{\rm sp}$ from the $R_{c}$ selection may not
be applicable directly to our data measurements.

To test the impact of $R_{c}$ on our splashback measurements, we re-run \redmapper, 
setting $R_{0}=0.75$ and 1.25 $h^{-1}{\rm Mpc}$. The resultant cluster catalogs will have 
a new richness estimate $\lambda'$ for each cluster. We rank in descending order the old 
cluster catalog by $\lambda$ and the new cluster catalog by $\lambda'$, then select the 
clusters in the new catalog that have the same ranking as 
the fiducial sample in our original cluster catalog with $20<\lambda<100$. We find 
$16.5<\lambda'<75.3$ ($22.8<\lambda'<117.3$) gives roughly the same number of 
clusters with the same ranking for $R_{0}=0.75$ (1.25) $h^{-1}{\rm Mpc}$. We measure 
$\Sigma_{g}$ and $\Delta \Sigma$ for these two new cluster catalogs and fit them to our model. 
We have checked that the amplitude of the $\Delta \Sigma$ measurements are nearly identical 
for the different $R_{0}$ settings, suggesting that the mean mass of the samples did not change 
significantly when we change $R_{0}$.  We note that the choices of $R_{0}$ here are rather 
extreme and the \redmapper code is not well tested at these $R_{0}$ values. For instance, we 
expect the scatter in the mass-richness relation to be much larger at these extreme $R_{0}$
values, which could have an effect on the resulting stacked profiles. As a result, the tests below 
should be treated as bounds for the potential systematic effects introduced by the $R_{c}$ 
settings.

\begin{figure}
\centering
\includegraphics[width=0.99\linewidth]{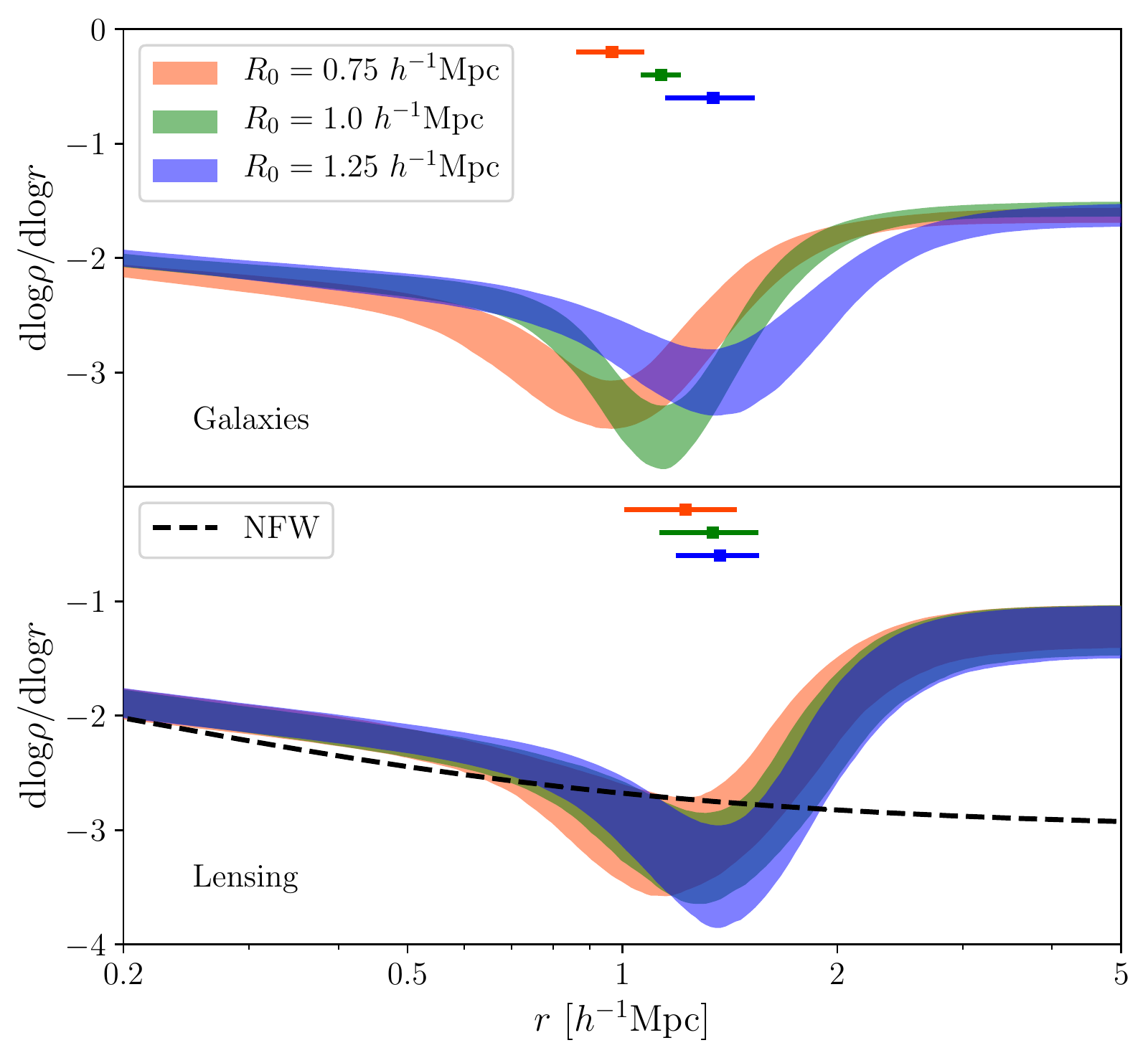}
\caption{Effect of changing $R_{0}$ on the inferred $r_{\rm sp}. $\textit{Top:} the log-derivative of 
the 3D model fits to the three $\Sigma_{g}$ measurements for the fiducial cluster sample of 
$20<\lambda<100$ with different $R_{0}$ settings during the \redmapper run.  $R_{0}=1h^{-1}$Mpc 
corresponds to the default \redmapper setting. \textit{Bottom:} same as the top panel but for weak 
lensing measurements. The dashed line shows the log-derivative of an NFW profile with a similar 
mass to these clusters.}
\label{fig:rc_test}
\end{figure}

The top panel of \Fref{fig:rc_test} shows the log-derivative of the
model fits to $\Sigma_{g}$ for these two cases together with the
fiducial setting of $R_{0}=1$ $h^{-1}{\rm Mpc}$. We find that the
profiles do indeed change as a function of $R_{0}$, similar to what
was seen in \citet{Busch2017}. The quantitative change in our
measurements is, however, smaller than that seen in \citet{Busch2017},
likely because \citet{Busch2017} employed a simplified \redmapper-like
cluster finder.

For $\Delta\Sigma$, we show in the lower panel of \Fref{fig:rc_test}
the resulting log-derivative of the model fits for the two alternative
$R_{0}$'s. Similar to what is seen in the $\Sigma_{g}$, the location
of $r_{\rm sp}$ moves outwards as the $R_{0}$ value
increases. However, for each of the $R_{0}$ settings, the
lensing-inferred $r_{\rm sp}$ remains consistent with the galaxy
measurements at better than $1\sigma$.  Also, compared to the NFW
profile shown by the dashed black curve, the slope at $r_{\rm sp}$ for
the lensing remains steeper than NFW by about 1$\sigma$. We note,
however, that the lensing-inferred $r_{\rm sp}$ appears to be more
robust to the change in $R_{0}$ than the galaxies.
 
The variation of $r_{\rm sp}$ with $R_0$ is not necessarily indicative
of a systematic error in the measured mass profile or the inferred values 
of $r_{\rm sp}$. Instead, it could suggest another source of selection effect 
in \redmapper. By changing $R_0$, one is selecting a new sample of clusters,
which could in principle have physically different $r_{\rm sp}$. One
might imagine, for instance, that changing $R_0$ could be analogous to
selecting clusters on $R_{\rm mem}$. If $R_0$ is
decreased, then we would expect to select clusters that have galaxies
that are more centrally concentrated, which would have smaller $R_{\rm  mem}$. 
In this case, we would expect to see $r_{\rm sp}$ change
with changing $R_0$ since we know that selecting clusters with
different $R_{\rm mem}$ leads to different inferred splashback radii.
Such selection effects will impact the comparison of the data
measurements to simulations, since \redmapper selection may not be
equivalent to the mass selection used in the simulations.

We expect the lensing measurements to be somewhat less affected by
\redmapper selection effects than galaxy density measurements since
the \redmapper selection is done directly on galaxies and not on the
shears.  This could also suggest that comparing the lensing
measurements with the dark matter simulations is a cleaner approach
and bypasses some of the \redmapper systematic issues.  To avoid such
selection effects altogether, one alternative is to use clusters
selected via X-ray or Sunyaev-Zel'dovich effect instead of optical
cluster finders.  However, such catalogs are typically smaller than
optically-selected catalogs, making high signal-to-noise measurements
difficult.

While the problems outlined above are certainly worrying, we note that our main findings of the 
analysis concern the detection of the splashback feature and the {\it relative} position of the 
splashback radius between the galaxy and lensing measurements, which we have shown above to 
be unaffected even when we use extreme vales of $R_{0}$. In \Aref{sec:rc_test} we show additional 
tests on the effect of changing $R_{0}$ on other analyses in this paper. 
The main comparison that makes use of the absolute value of $r_{\rm sp}$ in our analysis is the 
comparison of data measurements with the dark matter simulations in \Sref{sec:results_sims}. We 
therefore conclude that we cannot rule out the possibility that the apparent discrepancy between the 
galaxy and the dark matter-inferred $r_{\rm sp}$ results from such a selection effect. This needs to 
be quantified and understood more thoroughly before invoking physical explanations.

\section{Summary}
\label{sec:discussion}

The splashback feature has recently been pointed out as  a new probe for physics on the cluster 
scales. As the theory behind the splashback process is relatively clean, it can provide a physically 
motivated definition of the halo boundary, as well as a potential laboratory for tests of dark matter 
physics and gravity.

In this work we have measured the splashback feature around \redmapper clusters in the first year 
of DES data (DES Y1) using both the stacked galaxy density profile and the stacked weak lensing 
mass profiles. Our main analysis is based on a fiducial cluster sample of 3,684 clusters at redshift 
$0.2<z<0.55$ and richness $20<\lambda<100$. We apply the methodology developed in 
\citet{More2016} and \citet{Baxter2017} to DES Y1 data and expand the analysis in several aspects 
compared to previous work in SDSS.

We analyze the lensing measurements and demonstrate the existence of a splashback-like 
steepening in the outer mass profile of galaxy clusters. Furthermore, the location ($r_{\rm sp}$) and 
steepness of this truncation inferred from the mass profile agrees well with what is inferred from the 
stacked galaxy density measurements. The agreement in $r_{\rm sp}$ between galaxies and weak 
lensing is encouraging as it directly measures the mass distribution of the halo profiles. For measurements 
from the galaxy density (weak lensing) profiles, we constrain the cluster density profile at $r_{\rm sp}$ to be 
steeper than NFW at $3.0\sigma$ (2.0$\sigma$) significance when considering the total profile and 
4.6$\sigma$ (2.9$\sigma$) when considering only the collapsed material, which is the total profile 
subtracting out an infalling component. Future higher signal-to-noise lensing data will be able to test this 
statement with higher precision.

We compare our measurements to dark matter N-body simulations and find that, in 
agreement with previous results from SDSS, the $r_{\rm sp}$ measured from subhalos 
in simulations is higher than that measured with the galaxies. Compared to the lensing 
measurements, however, the discrepancy is only marginal due to the large uncertainty 
and slightly higher $r_{\rm sp}$ value in lensing. The level of the steepening is consistent 
with dark matter simulations. We also find differences in the overall shapes of the galaxy 
and lensing profiles compared with simulations. We note that selection effects in \redmapper 
can also affect these comparisons -- the clusters selected by the algorithm can have a slightly 
biased profile depending on the scale $R_{0}$ beyond which \redmapper cuts off member 
galaxies when estimating the richness.

We study the redshift and richness dependencies of $r_{\rm sp}$: we find no redshift evolution 
over the redshift range $0.3<z<0.6$ and a richness dependence consistent with expectation 
from $\Lambda$CDM simulations. However, the overall shape of the profiles for high and low 
richness clusters have some differences from what is measured in simulations.

Detection of dynamical friction is one of the applications of the splashback feature suggested 
by recent work \citep{Adhikari2016}. Massive galaxies falling into the potential of galaxy clusters 
will experience a drag force that is larger than the less massive galaxies, which would result in 
a smaller $r_{\rm sp}$. We measure the profile of galaxies in different luminosity bins around 
clusters and find that the highest luminosity galaxies indeed exhibit a slightly smaller $r_{\rm sp}$, 
a behavior that matches the corresponding subhalo profiles of the dark matter simulations. However, 
the difference is smaller than expected from simulation and within measurement uncertainties. We 
also tested that this measurement is sensitive to the change in $R_{0}$ mentioned above.

Looking towards the next DES data set which covers the full footprint of 5,000 deg$^{2}$, we can 
expect significant improvement in the statistical uncertainties in both the galaxy and the lensing 
measurements as well as the redshift coverage. However, interpreting the subtle systematic effects 
in the cluster finding algorithm and measurement process will be the crucial next step for a deeper 
understanding of the connection between the true splashback feature and the observed cluster 
profiles. One important step is to develop more realistic simulations that can reproduce the observables. 
In parallel, exploring the splashback feature for cluster samples selected in other wavelengths (in 
particular, SZ and X-ray selected samples) would be a good test for potential systematics in the optical 
cluster finder. A lensing mass selected cluster sample must await surveys that are deep enough to 
provide high significance detections of individual clusters. Improvements in simulations and the cluster 
selection will enable us to control for systematic effects and pursue the effects of standard and new 
physics associated with the splashback feature.

\section*{Acknowledgments}

CC and AK were supported in part by the Kavli Institute for
Cosmological Physics at the University of Chicago through grant NSF
PHY-1125897 and an endowment from Kavli Foundation and its founder
Fred Kavli. EB and BJ are partially supported by the US Department of
Energy grant DE-SC0007901. TNV was supported by the SFB-Transregio 33
\"The Dark Universe\" by the Deutsche Forschungsgemeinschaft (DFG) and
the DFG Cluster of Excellence "Origin and Structure of the
Universe". The weak lensing boost factors were calculated and
calibrated using the computing facilities of the Computational Center
for Particle and Astrophysics (C2PAP). DR is supported by a NASA Postdoctoral 
Program Senior Fellowship at the NASA Ames Research Center, administered by the 
Universities Space Research Association under contract with NASA.

The CosmoSim database used in this paper is a service by the
Leibniz-Institute for Astrophysics Potsdam (AIP).  The MultiDark
database was developed in cooperation with the Spanish MultiDark
Consolider Project CSD2009-00064.
The authors gratefully acknowledge the Gauss Centre for Supercomputing
e.V. (www.gauss-centre.eu) and the Partnership for Advanced
Supercomputing in Europe (PRACE, www.prace-ri.eu) for funding the
MultiDark simulation project by providing computing time on the GCS
Supercomputer SuperMUC at Leibniz Supercomputing Centre (LRZ,
www.lrz.de).  The Bolshoi simulations have been performed within the
Bolshoi project of the University of California High-Performance
AstroComputing Center (UC-HiPACC) and were run at the NASA Ames
Research Center.

Funding for the DES Projects has been provided by the U.S. Department of Energy, the U.S. National Science Foundation, the Ministry of Science and Education of Spain, 
the Science and Technology Facilities Council of the United Kingdom, the Higher Education Funding Council for England, the National Center for Supercomputing 
Applications at the University of Illinois at Urbana-Champaign, the Kavli Institute of Cosmological Physics at the University of Chicago, 
the Center for Cosmology and Astro-Particle Physics at the Ohio State University,
the Mitchell Institute for Fundamental Physics and Astronomy at Texas A\&M University, Financiadora de Estudos e Projetos, 
Funda{\c c}{\~a}o Carlos Chagas Filho de Amparo {\`a} Pesquisa do Estado do Rio de Janeiro, Conselho Nacional de Desenvolvimento Cient{\'i}fico e Tecnol{\'o}gico and 
the Minist{\'e}rio da Ci{\^e}ncia, Tecnologia e Inova{\c c}{\~a}o, the Deutsche Forschungsgemeinschaft and the Collaborating Institutions in the Dark Energy Survey. 

The Collaborating Institutions are Argonne National Laboratory, the University of California at Santa Cruz, the University of Cambridge, Centro de Investigaciones Energ{\'e}ticas, 
Medioambientales y Tecnol{\'o}gicas-Madrid, the University of Chicago, University College London, the DES-Brazil Consortium, the University of Edinburgh, 
the Eidgen{\"o}ssische Technische Hochschule (ETH) Z{\"u}rich, 
Fermi National Accelerator Laboratory, the University of Illinois at Urbana-Champaign, the Institut de Ci{\`e}ncies de l'Espai (IEEC/CSIC), 
the Institut de F{\'i}sica d'Altes Energies, Lawrence Berkeley National Laboratory, the Ludwig-Maximilians Universit{\"a}t M{\"u}nchen and the associated Excellence Cluster Universe, 
the University of Michigan, the National Optical Astronomy Observatory, the University of Nottingham, The Ohio State University, the University of Pennsylvania, the University of Portsmouth, 
SLAC National Accelerator Laboratory, Stanford University, the University of Sussex, Texas A\&M University, and the OzDES Membership Consortium.

Based in part on observations at Cerro Tololo Inter-American Observatory, National Optical Astronomy Observatory, which is operated by the Association of 
Universities for Research in Astronomy (AURA) under a cooperative agreement with the National Science Foundation.

The DES data management system is supported by the National Science Foundation under Grant Numbers AST-1138766 and AST-1536171.
The DES participants from Spanish institutions are partially supported by MINECO under grants AYA2015-71825, ESP2015-66861, FPA2015-68048, SEV-2016-0588, SEV-2016-0597, and MDM-2015-0509, 
some of which include ERDF funds from the European Union. IFAE is partially funded by the CERCA program of the Generalitat de Catalunya.
Research leading to these results has received funding from the European Research
Council under the European Union's Seventh Framework Program (FP7/2007-2013) including ERC grant agreements 240672, 291329, and 306478.
We  acknowledge support from the Australian Research Council Centre of Excellence for All-sky Astrophysics (CAASTRO), through project number CE110001020.

This manuscript has been authored by Fermi Research Alliance, LLC under Contract No. DE-AC02-07CH11359 with the U.S. Department of Energy, Office of Science, Office of High Energy Physics. The United States Government retains and the publisher, by accepting the article for publication, acknowledges that the United States Government retains a non-exclusive, paid-up, irrevocable, world-wide license to publish or reproduce the published form of this manuscript, or allow others to do so, for United States Government purposes.

This paper has gone through internal review by the DES collaboration.


\appendix

\section{Scaling measurements with richness $\lambda$}
\label{sec:lambda_scaling}

As shown in \citetalias{Diemer2014}, the splashback radius scales with
physical $R_{\rm 200m}$ and the accretion rate. We discussed in
\Sref{sec:gal_splashback} that our measurements are performed in
comoving distances which take into account the redshift evolution of
$R_{\rm 200m}$. Here we investigate the improvement in the measurements if
we were to also take into account the richness-dependence of
$R_{\rm 200m}$ within the sample. We do not attempt to correct for the
accretion rate-dependence in the same way, as the estimation of
accretion rates for clusters is non-trivial.

We sub-divide our fiducial cluster sample (with selection $20<\lambda<100$ and $0.2<z<0.55$) 
into 10 logarithmic $\lambda$ bins and repeat the measurement in 
\Sref{sec:gal_splashback}. The measured distances of galaxies from the cluster centers in 
each $\lambda$ bin are then scaled by $(\bar{\lambda}_i/\bar{\lambda}_{\rm full})^{F/3}$, where 
$\bar{\lambda}_i$ is the mean richness in the bin, $\bar{\lambda}_{\rm full}$ is 
the mean richness in the full sample, and $F=1.12$ is the exponent of the 
mass-richness relation derived in \citet{Melchior2016}.

\begin{figure}
\center
\includegraphics[width=0.5\linewidth]{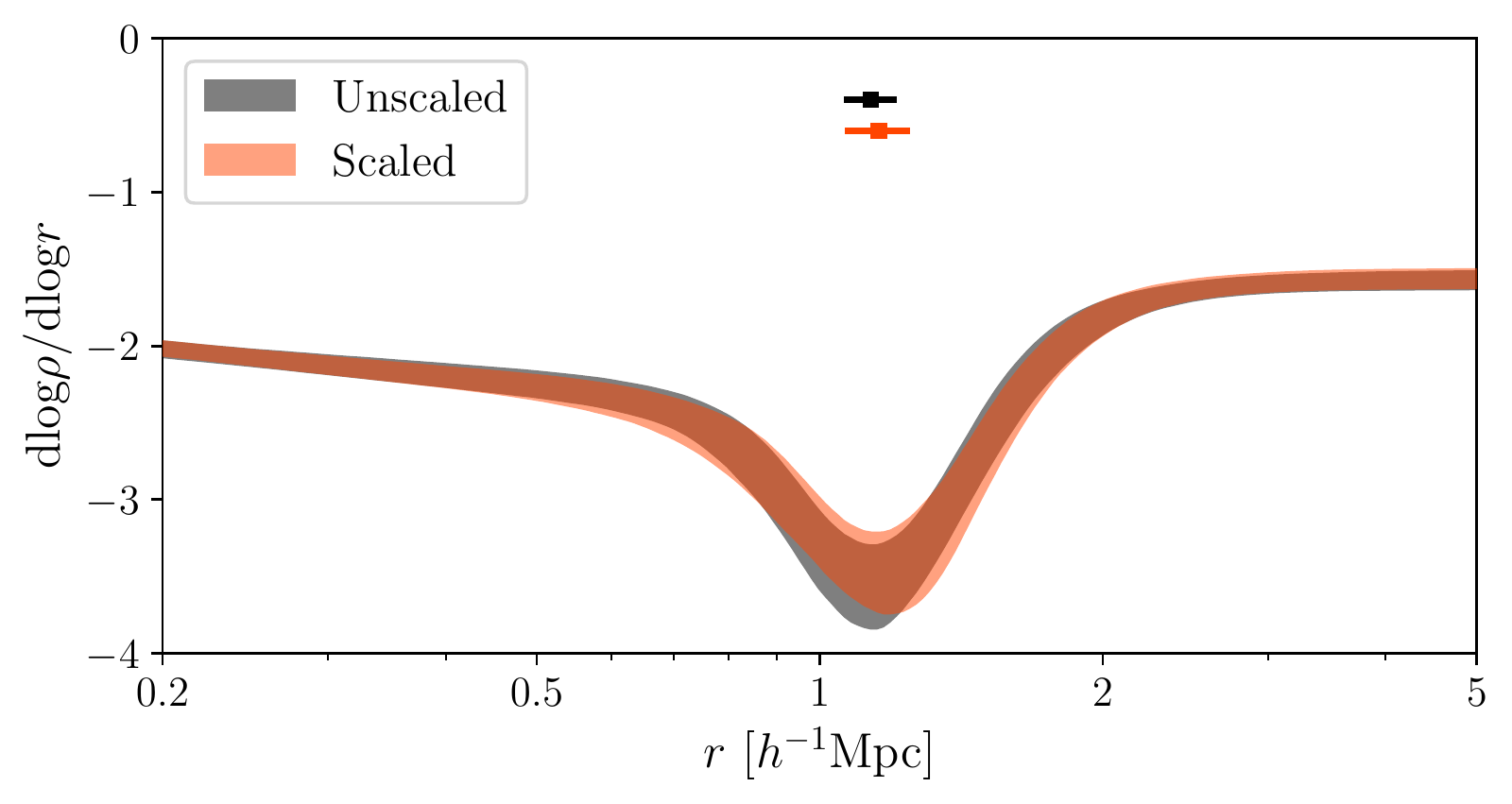}
\caption{Log-derivative for fiducial measurements with and without the 
$\lambda$-scaling that accounts for the range of $\lambda$ inside the bin. The 
inferred $r_{\rm sp}$ values are marked as horizontal bars on the top of the figure. 
The x-axis for the scaled case is $r/R_{\rm 200m}\bar{R}_{\rm 200m}$, where 
$\bar{R}_{\rm 200m}$ is the $R_{\rm 200m}$ at the mean richness.}
\label{fig:lambda_scaling}
\end{figure}

\Fref{fig:lambda_scaling} shows the log-derivative of the model fit to the fiducial 
measurement (\Fref{fig:splashback_fiducial}) including and not the 
$\lambda$-scaling. We find that after taking into account the $\lambda$ 
dependence of $r_{\rm sp}$ the splashback feature does not change
significantly. This is somewhat counterintuitive given the results in 
\Fref{fig: splashback_zdep_ldep}. Taking a closer look at the measurements, 
we find that the improvement in the $\lambda$-scaling is mostly washed out 
by the slightly increased error bars in the measurements, which is likely a 
result of the large scatter in the mass-richness relation. 

\section{Additional tests of the impact of $R_{0}$ on the splashback feature}
\label{sec:rc_test}

As discussed in \Sref{sec:model_sys}, we have seen that the choice of $R_{0}$ 
in \redmapper affects the inferred splashback radius, likely a result of selection effects. 
In this appendix, we carry out a few more tests to see the impact of $R_{0}$ on 
other measurements in this paper. 

\subsection{Effect of $R_{0}$ on $r_{\rm sp}$ in high richness clusters}

\begin{figure}
\centering
\includegraphics[width=0.5\linewidth]{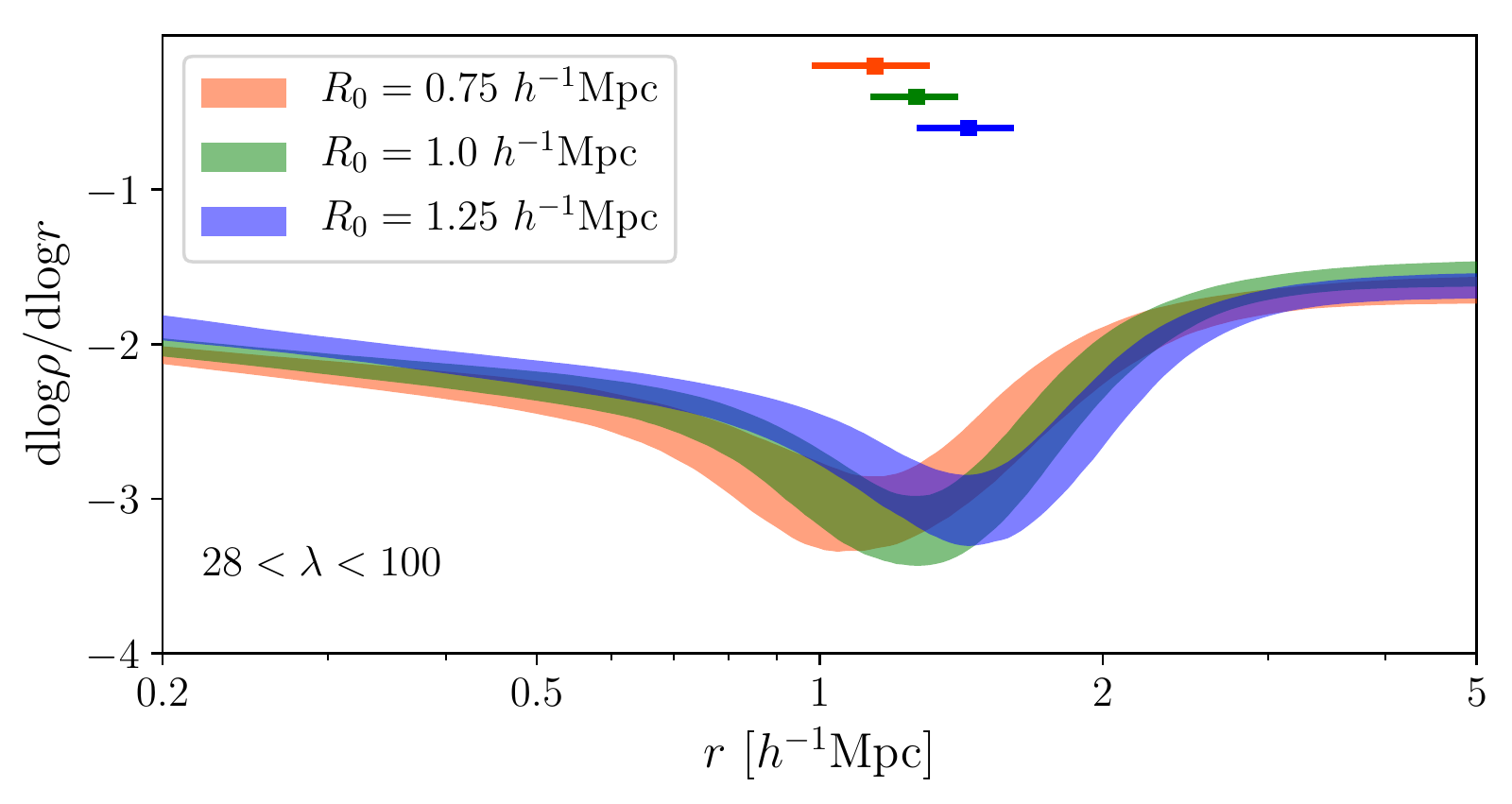}
\caption{Same as the top panel of \Fref{fig:rc_test} but for clusters of richness $28<\lambda<100$. }
\label{fig:rc_test_highlambda}
\end{figure}

Recently, using a combination of weak lensing and abundance measurements, 
\citet{Murata2017} found that the \redmapper mass-richness relation exhibits 
unexpectedly large scatter at low-richness. A non-negligible
fraction of the clusters with richness $\sim20$ come from halos of
mass $\approx 10^{13}$ $M_{\odot}$. One of the hypotheses in
\citet{Murata2017} is that the low-richness clusters are affected by
projection effects and thus less reliable. Inspired by this finding, we perform the test on the high-richness 
clusters in \Sref{sec:lamb_dep} to see whether they are more or less sensitive to the choice of $R_{0}$. 
The results are shown in \Fref{fig:rc_test_highlambda}. We find that the high-richness clusters 
give lower signal-to-noise results, and are similarly affected by the $R_{0}$ settings.  
We also compared the $r_{\rm sp}$ inferred from the high richness clusters with what is 
expected from the dark matter simulations, and do not see significantly improved 
agreement. These results show that with the statistical uncertainties in our 
data set, we do not gain by switching to a higher richness sample.

\subsection{Effect of $R_{0}$ on $r_{\rm sp}$ in dynamical friction measurements}

\begin{figure}
\center
\includegraphics[width=0.5\linewidth]{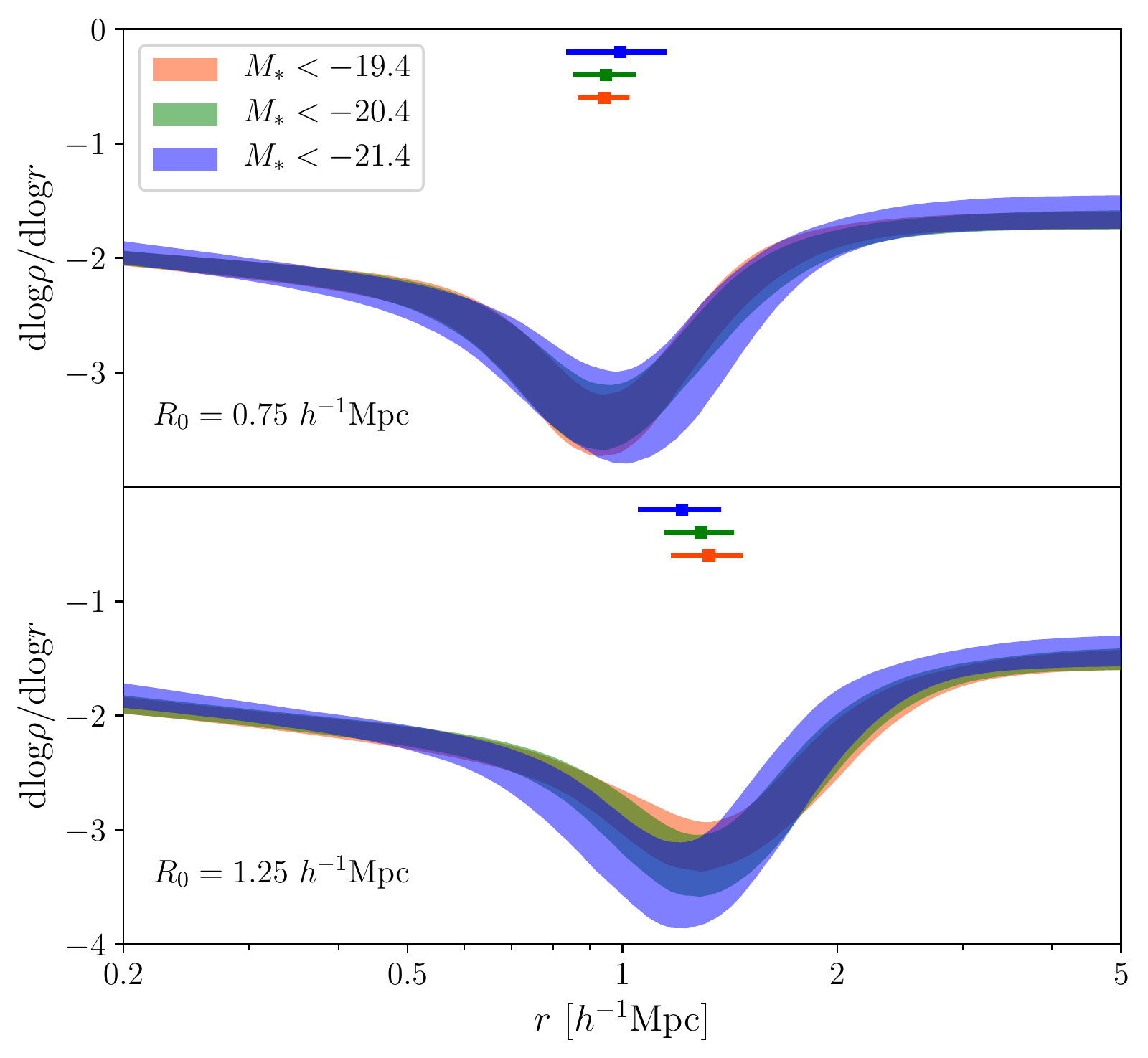}
\caption{Log-derivative for galaxy profiles of different luminosity, with cluster samples of different 
$R_{0}$ settings: the upper (lower) panel shows the same measurements as the bottom panel of 
\Fref{fig:dynamicalfriction_highlambda} but with $R_{0}=0.75$ ($R_{0}=1.25$) $h^{-1}$Mpc.}
\label{fig:rc_test_df}
\end{figure}

Here we test how the $R_{0}$ settings affect our measurements in \Sref{sec:df}. To do this, we repeat 
the measurements in \Sref{sec:df} using the two cluster catalogs described in \Sref{sec:model_sys}, 
which were derived using different $R_{0}$ values. The resulting measurements are shown in 
\Fref{fig:rc_test_df}. We find that with $R_{0}=1.25$ $h^{-1}$Mpc, the three galaxy luminosity bins 
show similar trends as our fiducial case of $R_{0}=1$ $h^{-1}$Mpc, where there is a hint of dynamical 
friction, but at lower significance. For the $R_{0}=0.75$ $h^{-1}$Mpc case the three galaxy sample gives 
consistent $r_{\rm sp}$ values and no sign of dynamical friction is seen. These findings again show that 
the measurement of $r_{\rm  sp}$ is sensitive to the choice of $R_{0}$.

\section{Effect of varying \lowercase{$h_{\rm max}$}}
\label{sec:hmax}

\begin{figure}
\center
\includegraphics[width=0.5\linewidth]{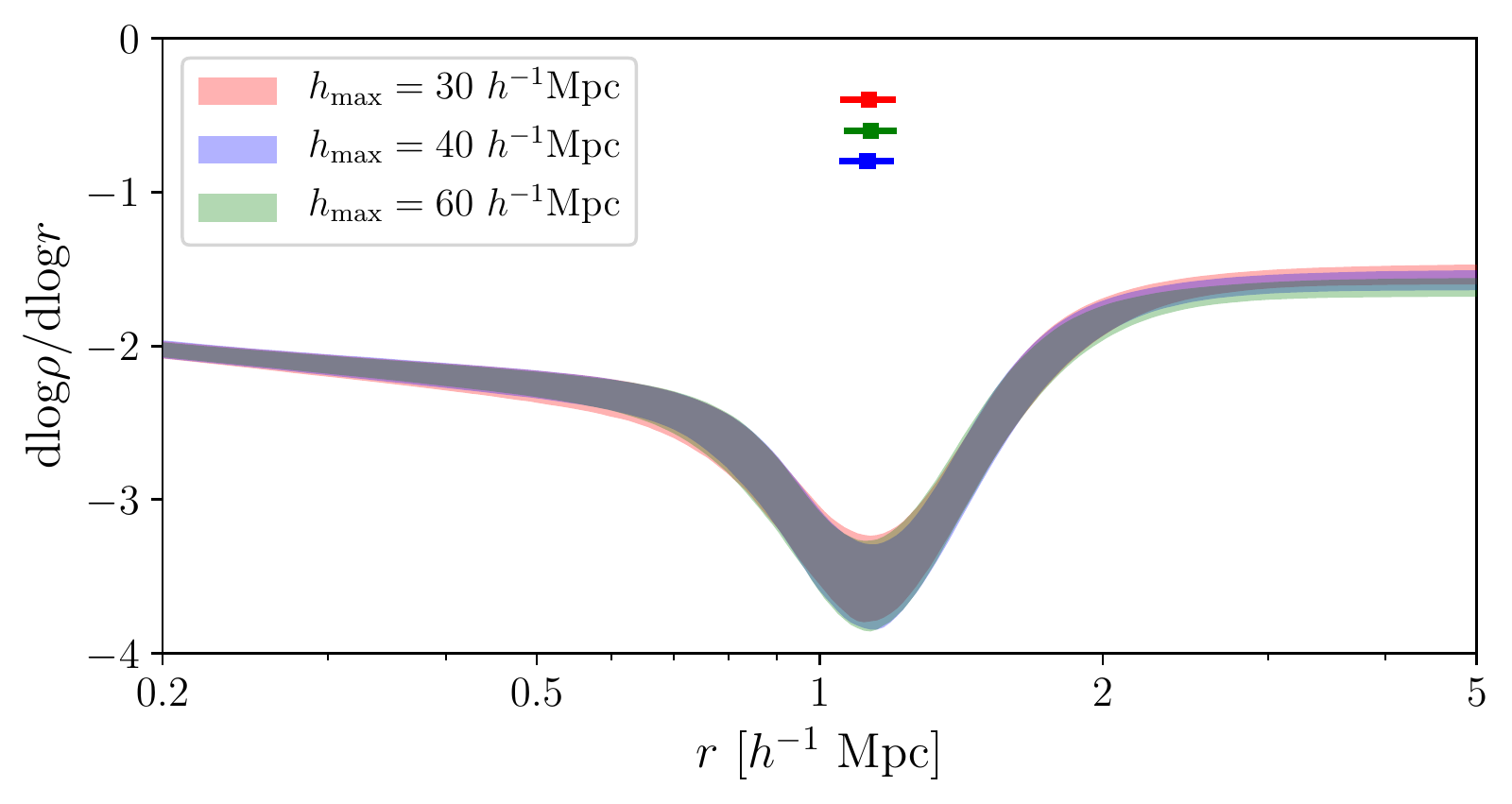}
\caption{Log-derivative of galaxy density profiles calculated from
  model fits for varying choices of $h_{\rm max}$, the maximum
  distance along the line of sight to integrate when converting the 3D
  profile into a projected profile. }
\label{fig:hmax_test}
\end{figure}

As discussed in \Sref{sec:model}, our model for the projected density profile,
$\Sigma(R)$ is obtained by integrating the 3D profile, $\rho(r)$ along
the line of sight.  Throughout this analysis, we impose a maximum line
of sight integration distance of $h_{\rm max} = 40$ $h^{-1}$Mpc.
Fig.~\ref{fig:hmax_test} shows the effect on our results of varying
$h_{\rm max}$.  In general, we find that our inferences about the 3D
profile are quite insensitive to the choice of $h_{\rm max}$.
There is a small change in the inferred slope of the outer density profile, but 
we note that most of our main results are not sensitive to the precise value 
this outer density profile.

\end{document}